\documentclass[apj]{emulateapj}

\shorttitle{Exoplanet Compositional Degeneracies}
\shortauthors{Rogers et al.}

\usepackage{graphicx}
\usepackage{epstopdf}
\usepackage{epsfig}
\usepackage{natbib}
\usepackage{wasysym}
\begin{document}

\title{A Framework for Quantifying the Degeneracies of Exoplanet Interior Compositions}

\author{L. A. Rogers\altaffilmark{1} and S. Seager \altaffilmark{1,} \altaffilmark{2}}

\altaffiltext{1}{Department of Physics, Massachusetts Institute of Technology, Cambridge, MA 02139, USA}
\altaffiltext{2}{Department of Earth, Atmospheric, and Planetary Sciences, Massachusetts Institute of Technology, Cambridge, MA 02139, USA}

\begin{abstract}

Several transiting super-Earths are expected to be discovered in the coming few years. While tools to model the interior structure of transiting planets exist, inferences about the composition are fraught with ambiguities. We present a framework to quantify how much we can robustly infer about super-Earth and Neptune-size exoplanet interiors from radius and mass measurements. We introduce quaternary diagrams  to illustrate the range of possible interior compositions for planets with four layers (iron core, silicate mantles, water layers, and H/He envelopes). We apply our model to CoRoT-7b, GJ~436b, and HAT-P-11b. Interpretation of planets with H/He envelopes is limited by the model uncertainty in the interior temperature, while for CoRoT-7b observational uncertainties dominate. We further find that our planet interior model sharpens the observational constraints on CoRoT-7b's mass and radius, assuming the planet does not contain significant amounts of water or gas. We show that the strength of the limits that can be placed on a super-Earth's composition depends on the planet's density; for similar observational uncertainties, high-density super-Mercuries allow the tightest composition constraints. Finally, we describe how techniques from Bayesian statistics can be used to take into account in a formal way the combined contributions of both theoretical and observational uncertainties to ambiguities in a planet's interior composition. On the whole, with only a mass and radius measurement an exact interior composition cannot be inferred for an exoplanet because the problem is highly underconstrained. Detailed quantitative ranges of plausible compositions, however, can be found.

\end{abstract}


\section{Introduction}

Over two dozen low-mass exoplanets with masses less than 30 Earth masses are known\footnote{See \url{exoplanet.eu} and references therein.}.  As their numbers increase, so does the probability to uncover a population of transiting low-mass exoplanets. The first transiting super-Earth exoplanet has been discovered \citep{LegerEt2009A&A}-based on the young history of exoplanets once one example of new type of object is discovered many more soon follow. Now that we are on the verge of discovering a good number of low-mass transiting planets \citep{BaglinEt2009IAUS, BoruckiEt2008IAUS, IrwinEt2009IAUS,  MayorEt2009A&A, LovisEt2009IAUS}, methods  to constrain their interior composition from observations are required. 

A good example of why quantitative methods to constrain planetary interior compositions are needed is GJ~436b \citep{ButlerEt2004ApJ, GillonEt2007A&Aa}, a Neptune-mass ($M_p=23.17\pm0.79 M_{\oplus}, $; \citet{TorresEt2008ApJ}), Neptune-size ($R_p=4.22^{+0.09}_{-0.10} R_{\oplus}$; \citet{TorresEt2008ApJ}) planet in a 2.6-day period around an M2.5 star. Initially, because of its similarity to the physical proportions of Neptune, \citet{GillonEt2007A&Aa} assumed that GJ~436b was composed mostly of ices. Others showed that it could instead be composed of a rocky interior with a more massive H/He envelope \citep{AdamsEt2008ApJ}.

Previously, \citet{ValenciaEt2007bApJ} introduced ternary diagrams to constrain the interior composition of  super-Earths without gas envelopes. \citet{Zeng&Seager2008PASP} presented a detailed description of the functional form of the ternary diagram interior composition curves. Super-Earths are loosely defined as planets with masses between 1 and 10 Earth masses that are composed of rocky or iron material. While the terms ``mini-Neptune" or ``Neptune-like" are not in common usage, they refer to planets with significant gas envelopes. Others have modeled evolution of Neptune-mass planets to predict radii \citep[e.g.][]{FortneyEt2007ApJ, BaraffeEt2006A&A}. \citet{FigueiraEt2009A&A} used planet formation and migration models to suggest interior compositions for GJ~436b.

In this paper, we aim to quantify the constraints placed on a low-mass exoplanet's interior structure by transit and radial velocity observations. We use a planetary structure model to explore the range of plausible interior compositions that are consistent with a given pair of mass and radius measurements, independent of planet formation arguments. We extend previous work by including the possibility of a gas envelope and by considering a range of mantle iron enrichments. Our model of low-mass planet interiors includes an iron core, silicate mantle, water ice layer, and H/He layer. To plot four-layer interior compositions we introduce quaternary diagrams, an expansion of ternary diagrams into three dimensions. Finally, we present a new framework to combine both model and observational uncertainties in a rigorous way using Bayesian techniques when interpreting the interior composition of a transiting exoplanet. Our overall goal is to be able to interpret planetary mass and radius observations with a quantitative understanding of the effects of model uncertainties, observational uncertainties, and the inherent degeneracy originating from the fact that planets of differing compositions can have identical masses and radii. 

We describe our planetary interior structure model in \S~\ref{sec:model}. We introduce quaternary diagrams in \S~\ref{sec:Quat}. In \S~\ref{sec:results}, we apply our model to constrain the compositions of low-mass exoplanets. In \S~\ref{sec:Bayesian}, we describe how Bayesian techniques may be applied to the problem of drawing inferences about an exoplanet's interior from measurements of the planet's mass and radius. Discussion and conclusions follow in \S~\ref{sec:Dis} and \ref{sec:con}. 

\section{Model}
\label{sec:model}

\subsection{Model Overview}
\label{sec:mo}

We consider a spherically symmetric differentiated planet in hydrostatic equilibrium. With these assumptions, the radius $r\left(m\right)$ and pressure $P\left(m\right)$, viewed as functions of the interior mass $m$, obey the coupled differential equations 
	
\begin{eqnarray}
\frac{dr}{dm}&=&\frac{1}{4\pi r^2\rho}\label{eqn:dr},\\
\frac{dP}{dm}&=&-\frac{Gm}{4\pi r^4}\label{eqn:dP},
\end{eqnarray}

\noindent where $\rho$ is the density and $G$ is the gravitational constant. Equation~(\ref{eqn:dr}) is derived from the mass of a spherical shell, while Equation~(\ref{eqn:dP}) describes the condition for hydrostatic equilibrium. The equation of state (EOS) of the material 

\begin{equation}
\rho = f\left(P, T\right)
\end{equation}

\noindent relates the density $\rho\left(m\right)$ to the pressure $P\left(m\right)$ and temperature $T\left(m\right)$ within a layer. We allow our model planets to have several distinct chemical layers ordered such that the density $\rho\left(m\right)$ is monotonically decreasing as $m$ increases toward the planet surface. Throughout the rest of this work, we shall use $x_i=M_i/M_p$ to denote the fraction of a planet's total mass $M_p$ in the $i$th layer from the planet center ($i=1$ denotes the innermost layer). 

	To model a planet having mass $M_p$, radius $R_p$, and a specified composition $\{x_i\}$, we employ a fourth-order Runge-Kutta routine to numerically integrate Equations~(\ref{eqn:dr}) and~(\ref{eqn:dP}) for $r\left(m\right)$ and $P\left(m\right)$ from the outer boundary of the planet $\left(m=M_p\right)$ toward the planet center  $\left(m=0\right)$. We describe our scheme for setting the exterior boundary conditions in \S~\ref{sec:bc}. We impose that both $P$ and $r$ are continuous across layer boundaries. At each step in the integration, the EOSs and temperature profiles described in \S~\ref{sec:EOS} are used to evaluate $\rho\left(m\right)$.
	
	The planet parameters $\{M_p,\,R_p,\,\{x_i\}\}$ in fact form an overdetermined system; there is a single radius $R_p$ that is consistent with $M_p$ and $\{x_i\}$. For a given mass and composition, we use a bisection root-finding algorithm to iteratively solve for the planet radius $R_p$ that yields $r\left(m=0\right)=0$ upon integrating Equations~(\ref{eqn:dr}) and (\ref{eqn:dP}) to the planet center. We stop the iteration once we have found $R_p$ to within $100~\rm{m}$. Alternatively, in some applications it is convenient to be able to stipulate the planet radius (for instance when exploring the range of compositions $\{x_i\}$ allowed for a confirmed transiting exoplanet of measured mass and radius). In these cases, we use a bisection root-finding algorithm to iteratively solve for the mass ratio of the inner two material layers $\left(x_2/x_1\right)$ of the planet given $M_p$, $R_p$, and valid mass distribution in the outer layers of the planet  $\{x_i\,|\,i>2\}$. We stop this iteration once we have found $x_1$ and $x_2$ to within $10^{-10}$.
		
	We increase the achievable accuracy in the composition of our modeled planets and the stability of this iterative process by employing the Lagrangian form of the equations of structure. With mass $m$ as the independent integration parameter, we can take a partial mass step at the conclusion of each layer  $i$ to ensure that the specified value of $x_i$ is precisely obtained. Within each layer, we employ an adaptive mass step-size such that each integration step corresponds to a radius increment of approximately $100~\rm{m}$. An adaptive step-size is necessary because both Equations~(\ref{eqn:dr}) and (\ref{eqn:dP}) diverge as $r\to0$ and $m\to0$.

\subsection{Material EOS and Thermal Profiles}
\label{sec:EOS}

	In this section, we describe the EOS and thermal profile $T\left(m\right)$ assumed for each material layer. 

	We allow for the presence of an outer gas envelope in our modeled planets. We use the H/He EOS with helium mass fraction $Y=0.28$ from \citet{SaumonEt1995ApJS}. As mentioned in \citet{AdamsEt2008ApJ}, we ignore the ``plasma phase transition" in the H/He EOS.  To set the thermal profile we divide the H/He layer into two regimes: a thin outer radiative layer and an inner convective layer. 
	
	In the radiative regime of the gas envelope, we employ the analytic work of \citet{Hansen2008ApJS} to approximate the temperature profile.  \citet{Hansen2008ApJS} considered a plane-parallel atmosphere in radiative equilibrium that is releasing heat flux generated in the planet interior while also being irradiated by a mono-directional beam of starlight. He solved the gray equations of radiative transfer with a `two-stream' approach, allowing the incoming optical stellar photons to have a different opacity and optical depth than the infrared photons reradiated by the planet, and obtained a temperature profile 

\begin{eqnarray}
T^4&=&\frac{3}{4}{T_{\rm{eff}}}^4\left[\tau+\frac{2}{3}\right]+\mu_0{T_0}^4\left[1+\frac{3}{2}\left(\frac{\mu_0}{\gamma}\right)^2\right.\nonumber\\
&&\left.-\frac{3}{2}\left(\frac{\mu_0}{\gamma}\right)^3\mathrm{ln}\left(1+\frac{\gamma}{\mu_0}\right)-\frac{3}{4}\frac{\mu_0}{\gamma}e^{-\gamma\tau/\mu_0}\right].
\label{eq:Trad}
\end{eqnarray}

\noindent In the above equation, $T$ is the atmospheric temperature, $\tau$ is the infrared optical depth, $\gamma$ is the ratio between the optical and infrared optical depths, $\mu_0$ is angle cosine of the incoming beam of starlight relative to the local surface normal, $T_{\rm{eff}}$ is the effective temperature of the planet in the absence of stellar irradiation, and $T_{0}$ characterizes the magnitude of the stellar flux at the orbital distance of the planet $\left(F_*\left(R_*/a\right)^2=\sigma {T_0}^4\right)$. While $\mu_0$ varies over the surface of the planet, our planet model is one-dimensional spherically symmetric model. We adopt a single fiducial value of $\mu_0=1/2$ (the average of $\mu_0$ over the day hemisphere) when calculating the temperature profile of the radiative gas layer.  Equation~(\ref{eq:Trad}) yields the temperature in the radiative gas layer as a function of the (infrared) optical depth. The variation of optical depth $\tau$, with interior mass $m$ obeys

\begin{equation}
\frac{d\tau}{dm}=-\frac{\kappa}{4\pi r^2},
\label{eqn:dtau}
\end{equation}
 
\noindent where $\kappa$ is the opacity. In the radiative regime of the gas layer, we integrate Equation~(\ref{eqn:dtau}) along with Equations~(\ref{eqn:dr}) and (\ref{eqn:dP}). For $\kappa$,  we use tabulated Rosseland mean opacities of H/He at solar abundance metallicity $\left(\left[M/H\right]=0.0\right)$ from \citet{FreedmanEt2008ApJS}.

In our model gas layer, we allow for the presence of an inner adiabatic regime within which energy transport is dominated by convection.  Neglecting the effects of conduction and diffusion, we take the temperature profile in the convective layer to follow the adiabat fixed to the specific entropy at the base of the radiative regime. The transition between the radiative and convective regimes is determined by the onset of convective instabilities. An adiabatically displaced fluid element in the gas layer will experience a buoyancy force tending to increase its displacement if

\begin{equation}
0<\left(\frac{\partial \rho}{\partial s}\right)_P\frac{ds}{dm}=-\frac{\rho}{V}\left(\frac{\partial T}{\partial P}\right)_s\frac{ds}{dm},
\label{eqn:radCon}
\end{equation}

\noindent where the density $\rho\equiv\rho\left(P,s\right)$ is viewed as a function of the pressure $P$ and specific entropy per unit mass $s$. Whenever Equation~(\ref{eqn:radCon}) is satisfied, the gas layer is unstable to convection. In the H/He EOS from \citet{SaumonEt1995ApJS}, the adiabatic gradient $\left(\partial T/\partial P\right)_s$ is positive for all values of $P$, $T$, and He mass fraction $Y$. It thus suffices to test for $ds/dm<0$ to define the outer boundary of the convection regime. As we integrate Equations~(\ref{eqn:dr}), (\ref{eqn:dP}), and (\ref{eqn:dtau}) from the planet exterior inward, we transition from the radiative regime to the convective regime once  $ds/dm<0$.

	In the interior solid layers of the planet, we neglect the temperature dependence of the EOS. Thermal effects in the solid layers of a planet have a small effect on the planet radius \citep{SeagerEt2007ApJ} justifying the assumption of a simplified isothermal temperature profile. For every solid material considered in this study (Fe, FeS, $\rm{Mg_{1-\chi}Fe_{\chi}SiO_3}$, and $\rm{H_2O}$) we use the EOS data sets from \citet{SeagerEt2007ApJ} derived by combining experimental data at $P\lesssim200~\rm{GPa}$ with the theoretical Thomas-Fermi-Dirac EOS at high pressures, $P\gtrsim10^4~\rm{GPa}$.

\subsection{Exterior Boundary Condition}
\label{sec:bc}

	In our model (described in \S~\ref{sec:mo}), the exterior boundary of the planet sets the initial conditions for integrating the equations of structure. In the absence of a gas layer, we take the pressure to be $0$ at the solid surface of the planet $\left(m=M_p,\,r=R_p,\,P=0\right)$. For planets having gas layers, we use a simplified constant scale height atmospheric model to choose appropriate exterior boundary conditions on the pressure $P$ and optical depth $\tau$ at  $r\left(M_p\right)=R_p$ as elaborated below.  
	
	To physically motivate our choice of exterior boundary conditions for planets with gas layers, we make several simplifying approximations about the properties of the planet gas layer in the neighborhood of the measured planet radius $R_p$. We assume that in this region the gas layer can be approximated as an ideal gas, so that
	
\begin{equation}
P =\frac{\rho k_BT}{\mu_{\rm{eff}}},
\end{equation}

\noindent where $\mu_{\rm{eff}}$ is the effective molecular mass of the gas. We further assume that the outer atmosphere of the planet is isothermal. This is consistent with the radiative temperature profile from \citet{Hansen2008ApJS} (Equation~(\ref{eq:Trad})), which is largely isothermal for $\tau\ll1$. We also neglect variations in the surface gravity $g=GM/R^2$ over the range of radii being considered. Finally, to account for the pressure dependence of the opacity, we assume a power-law dependence

\begin{equation}
\kappa =CP^{\alpha}T^{\beta},
\end{equation}

\noindent where $\log{C}=-7.32$, $\alpha=0.68$, and $\beta=0.45$ are determined by fitting to the \citet{FreedmanEt2008ApJS} tabulated opacities (with all quantities in SI units). These assumptions, when coupled with the equation of hydrostatic equilibrium $\left(dP/dr=-\rho g\right)$ and the definition of the radial optical depth $\left(d\tau/dr=-\kappa\rho\right)$, yield an exponential dependence of both $P$ and $\tau$ on $r$,

\begin{eqnarray}
P\left(r\right) &=& P_Re^{-\left(r-R_p\right)/H_P},\\
\tau\left(r\right)&=&\tau_Re^{-\left(\alpha+1\right)\left(r-R_p\right)/H_P},
\end{eqnarray}
	
\noindent with the pressure scale height $H_P$ given by

\begin{equation}
H_P =\frac{R_p^2k_BT}{GM_p\mu_{\rm{eff}}},
\end{equation}

\noindent and the pressure and optical depth at $R_p$ ($P_R$ and $\tau_R$, respectively) related by

\begin{equation}
P_R =  \left(\frac{GM_p\left(\alpha+1\right)\tau_R}{R_p^2CT^{\beta}}\right)^{1/\left(\alpha+1\right)}.
\label{eq:PR}
\end{equation}

	It is important to maintain a direct correspondence to observations when defining the radius of a gas-laden planet in our model. Planet radii are measured observationally from transit depths and thus reflect the effective occulting area of the planet disk. We denote the optical depth for absorption of starlight through the limb of the planet  $\tau_t\left(y\right)$, where $y$ is the cylindrical radius from the line of sight to the planet center. In our models we define the transit radius $R_p$ to occur at 
	
\begin{equation}
\tau_t\left(R_p\right)=1.
\label{eq:Rdef}
\end{equation}	

\noindent We use a development similar to that in  \citet{Hansen2008ApJS} to relate the transverse optical depth through the limb to the radial optical depth $\tau$. Integrating along the line of sight through the planet limb, the transverse optical depth for starlight is given by: 

\begin{eqnarray}
\tau_t\left(y\right) &=& 2\gamma\int_y^{\infty}\frac{\kappa\left(r\right)\rho\left(r\right)}{\sqrt{1-\left(y/r\right)^2}}dr\nonumber\\
& \approx&  \gamma\tau_R\sqrt{\frac{2\pi\left(\alpha+1\right)y}{H_P}}e^{-\left(\alpha+1\right)\left(y-R_p\right)/H_P}.
\label{eqn:taut4}  
\end{eqnarray}

\noindent The right-hand side of Equation~(\ref{eqn:taut4}) is obtained by recognizing that for $y\sim R_p\gg H_P/\left(\alpha+1\right)$ only values of $r$ with $\left(r-y\right)\ll y$ contribute significantly to the integral due to the exponential decay of the integrand. We obtain exterior boundary condition on $\tau$ by combining our model definition of the transit radius (Equation.~(\ref{eq:Rdef})) with Equation~(\ref{eqn:taut4}):

\begin{equation}
\tau_R=\frac{1}{\gamma}\sqrt{\frac{H_P}{2\pi\left(\alpha+1\right)R_p}}. 
\label{eq:tauR}
\end{equation}

\noindent The boundary condition for pressure follows from $\tau_R$ using Equation~(\ref{eq:PR}).

\subsection{Model Parameter Space}
\label{sec:Ebudget}
  
    In this section, we describe our procedure for choosing value ranges for $\gamma$, $T_0$, and $T_{\rm{eff}}$ that describe the atmospheric absorption, stellar insolation, and the intrinsic luminosity of exoplanets simulated with our model. 
  
  The parameter $\gamma$ in Equation~(\ref{eq:Trad}) denotes the ratio of the gas layer's optical depth to incident starlight over its optical depth to thermal radiation. At large values of $\gamma$ the starlight is absorbed high in the atmosphere, while at small values of $\gamma$ the stellar energy penetrates deeper into the atmosphere. We adopt a fiducial value of $\gamma=1$, but also consider values spanning from 0.1 to 10. In this way, we encompass a wide range of possible absorptive properties in our model H/He envelopes.

   In Equation~(\ref{eq:Trad}), $\mu_0\sigma{T_0}^4$ represents the stellar energy flux absorbed (and reradiated) locally at a given point on the planet's irradiated hemisphere. The stellar insolation impinging on a planet can be calculated with knowledge of the host star's luminosity $L_*$ or spectral class, and of the semimajor axis $a$ of the planet's orbit. The fraction of this energy that is reflected by the planet and how the energy that does get absorbed is distributed around the planet's surface area, however, remain unknown for the super-Earth and hot Neptune planets considered in this paper. Our parameterization of the energy received at the planet from the star is further complicated by the fact that we are using a spherically symmetric planetary model, whereas the effect of stellar insolation varies over the planet surface. We take these uncertainties into account by considering a range of plausible $T_0$ values for each planet. For our fiducial value, we use the equilibrium temperature of the planet assuming full redistribution and neglecting reflection
   
\begin{equation}
T_{0}=\left(\frac{L_*}{16\pi \sigma a^2}\right)^{1/4}.
\end{equation}   

\noindent Similar fiducial choices of $T_0$ have been made in other studies that used Equation~(\ref{eq:Trad}) to describe the gas layer temperature profiles of low-mass exoplanets \citep{AdamsEt2008ApJ, MillerRicciEt2009ApJ}. By considering reflection of starlight by the planet in addition to full redistribution, we set a lower bound on $T_0$: 

\begin{equation}
T_{0}=\left(\frac{L_*\left(1-A\right)}{16\pi \sigma a^2}\right)^{1/4},
\label{eq:TeqA}
\end{equation}  

\noindent where $A$ is the planet's Bond albedo. A plausible upper limit of $A=0.35$ is chosen; all the solar system planets except Venus have Bond albedos below this value. Finally, to establish an upper limit on $T_0$ we neglect both redistribution and reflection and take

\begin{equation}
T_{0}=\left(\frac{L_*}{4\pi \sigma a^2}\right)^{1/4}.
\end{equation} 

\noindent This upper bound corresponds to the formal definition of $T_0$ used by \citet{Hansen2008ApJS} to derive Equation~(\ref{eq:Trad}). 

	A planet's intrinsic luminosity (produced by radiogenic heating and by contraction and cooling after formation) is another important component of the planetary energy budgets. In Equation~(\ref{eq:Trad}), $T_{\rm{eff}}$ parameterizes the heat flux from the planet interior entering the planet gas layer from below, $F_{int}=\sigma{T_{\rm{eff}}}^4$. Within the plane-parallel gas layer assumption, we can relate $T_{\rm{eff}}$ to the intrinsic luminosity, $L_{int}$, of the planet
	
 \begin{equation}
T_{\rm{eff}}=\left(\frac{L_{int}}{4\pi \sigma R^2}\right)^{1/4}.
\end{equation} 

\noindent We require a scheme to constrain the intrinsic luminosities of low-mass exoplanets. 

A full evolution calculation, modeling the energy output of a planet as it ages after formation, is outside of the scope of this work. There are many physical effects (including phase separation, chemical differentiation, chemical inhomogeneities, irradiation, radiogenic heating, impacts, geological activity, tidal heating, and evaporation) that can influence the thermal evolution of a planet and flummox attempts to predict a planet's intrinsic luminosity (see \S~\ref{sec:disevolution} for a full discussion). Additionally, the ages of the planet-hosting stars considered here (and of the planets that surround them) are very poorly constrained. This severely limits the insights  that a cooling simulation could yield into the planets' intrinsic luminosities. Instead of directly simulating planetary evolution, we take an approximate scaling approach to bracket plausible values for the intrinsic luminosities of low-mass exoplanets. 

We use planet evolution tracks modeled by \citet{BaraffeEt2008A&A} to constrain the intrinsic luminosities of the gas-laden planets considered in this work. \citet{BaraffeEt2008A&A} modeled the evolution of planets ranging from $10~M_{\oplus}$ to $10~M_{\jupiter}$, having heavy metal enrichments of $Z= 2\%$, 10\%, 50\%, and 90\%, and that were either receiving negligible stellar irradiation or suffering insolation equivalent to that from a sun at 0.045~AU. We limit our consideration to the simulated irradiated planets that are at least 1 Gyr old and that are no more than $1~M_{\jupiter}$. We then fit the intrinsic luminosities of this sub-sample of  \citet{BaraffeEt2008A&A} models to a simple power law in planetary mass, radius, and age $\left(t_p\right)$:

 \begin{eqnarray}
\log{\left(\frac{L_{int}}{L_{\odot}}\right)}&=& a_1 + a_{M_p}\log{\left(\frac{M_{p}}{M_{\oplus}}\right)} + a_{R_p}\log{\left(\frac{R_{p}}{R_{\jupiter}}\right)}\nonumber\\
&& + a_{t_p}\log{\left(\frac{t_{p}}{\rm{1~Gyr}}\right)}. 
\label{eq:Lint}
\end{eqnarray} 

\noindent The values obtained for the coefficients and their 95\% confidence intervals are $a_1=-12.46\pm0.05$, $a_{M_p}=1.74\pm0.03$, $a_{R_p}=-0.94\pm0.09$, and $a_{t_p}=-1.04\pm0.04$. The fit had $R^2=0.978$ and rms residuals of 0.14 in $\log{\left(L_{int}/L_{\odot}\right)}$. For a given planet, we use the measured planetary mass, planetary radius, and host star age (a proxy for the planet age) with the  best-fit coefficients in Equation~(\ref{eq:Lint}) to calculate a fiducial value for $L_{int}$. We then employ the uncertainties in the fit coefficients, the rms residuals, and the range of possible planet ages to establish a nominal range of intrinsic luminosities $L_{int}$ to consider when constraining the interior compositions of planets with gas layers. The poorly constrained planet age dominates the other sources of uncertainties in its contribution to the range of $L_{int}$ for all the planets we consider.

Additional limitations on $T_{\rm{eff}}$ can be required if the nominal range of $L_{int}$ determined by the procedure above is too broad.  At very low values of $T_{\rm{eff}}$  (low intrinsic luminosities) the gas layer \textit{P-T} profile can enter an unphysical high-pressure  low-temperature regime $\left(P\gtrsim2.5\times10^{10}~\rm{Pa},\;T \lesssim 3500~\rm{K}\right)$. These conditions, under which hydrogen may form a Coulomb lattice or a molecular solid, are not included in the coverage of the \citet{SaumonEt1995ApJS} hydrogen EOS. If necessary, we truncate the lower range of $T_{\rm{eff}}$ values that we consider to avoid exceeding the range of applicability of the \citet{SaumonEt1995ApJS} EOS. Out of all the planets considered in the work, such a reduction in the range of $T_{\rm{eff}}$ was only required for HAT-P-11b (\S~\ref{sec:HATP11b}).

Adopting a simple scaling approach to estimate $T_{\rm{eff}}$ allows us to consider a wider variety of possible interior compositions than we could by simulating full evolution tracks. Nonetheless, our procedure to constrain $T_{\rm{eff}}$ is very approximate. It estimates the intrinsic luminosity of a planet from its mass, radius, and age alone. The effects of interior composition and stellar irradiation on a planet's evolution are not addressed. For instance, because solar system planets are less strongly irradiated than the transiting planets considered in this work, the scaling relation systematically overestimates their intrinsic luminosities. Further, the extrapolation of the \citet{BaraffeEt2008A&A} models to super-Earth-sized planets is very uncertain. Although phenomenological, the procedure described above provides a consistent way to estimate a plausible range of intrinsic luminosities  in which the span of  the range reflects the uncertainties in the planet age and thermal history.
	
\subsection{Model validation}

We have validated our planet interior model by comparing our results with Earth and other models. Our fiducial Earth-planet composition is one with a 32.6\% by mass core consisting of FeS (90\% iron and 10\% sulfur by mass) and a 67.4\% by mass mantle consisting of $\rm{Mg_{0.9}Fe_{0.1}SiO_3}$. For this composition, our model gives a radius of $6241~\rm{km}$ for a 1 $M_{\oplus}$ planet. This radius value is within 2.2\% of Earth's true radius, well within expected observational uncertainties for future discovered Earth-mass, Earth-sized exoplanets. More importantly, our solid planet models are not intended to be accurate for such low masses \citep{SeagerEt2007ApJ}, because we ignore thermal pressure. This approximation is much more appropriate for more massive planets, where a larger fraction of the planet's material is at high pressure where thermal effects are small.

We further compared our model output with the results presented in \citet{ValenciaEt2007ApJ}. Specifically, we reproduced the values in their Table 3 for GJ 876d's radius under various assumed bulk compositions . We found that for solid planets composed of iron and silicates our radii matched those from \citet{ValenciaEt2007ApJ} to 0.2\%. For planets that also included a water layer, our radii were within 1\%. This is a very reasonable agreement. The larger discrepancy in the water planet radii as compared to the dry-planet radii stems from differences in the EOS for water. 
See \citet{SeagerEt2007ApJ} for our calculations on the water EOS, and a detailed description of our EOS choices. 

We tested our model of planets with significant gas envelopes by comparing to \citet{BaraffeEt2008A&A} models of hot Neptunes. For planets of 10 and 20 $M_{\oplus}$ with 10\% by mass layer of H$_2$ and He, we found very good agreement between the model radii. The \citet{BaraffeEt2008A&A} radii fall within the range of planetary radii derived from our model when uncertainties on the atmospheric thermal profile and energy budget in our model are taken into account. In other words, it is possible to choose values of $T_{\rm{eff}}$, $\gamma$, and $T_0$ within the ranges described in \S~\ref{sec:Ebudget} such that our model radii agree precisely with those from \citet{BaraffeEt2008A&A}. Further, over the full range of atmospheric parameters considered our model radii deviate by no more than 27\% from those of \citet{BaraffeEt2008A&A}. 

\section{Ternary and Quaternary Diagrams}
\label{sec:Quat}

	In this work, we use ternary and quaternary diagrams to plot the relative contributions of the core, mantle, ice layer, and gas layer to the structure of a differentiated exoplanet. \citet{ValenciaEt2007bApJ}  and \citet{Zeng&Seager2008PASP} also employed ternary diagrams to present the interior composition of terrestrial exoplanets, and provide detailed discussions of these three-axis equilateral triangle diagrams. While both \citet{ValenciaEt2007bApJ}   and  \citet{Zeng&Seager2008PASP} considered three-component planets comprised of a core, a mantle, and water ices, our fiducial model also allows for a gas layer. Three-dimensional tetrahedron quaternary diagrams provide a natural extension of ternary diagrams to four-component systems.

	Quaternary diagrams are useful for plotting four-component data $\left(w, x, y, z\right)$ that are constrained to have a constant sum $\left(w+x+y+z=A=\rm{constant}\right)$. The axes of a quaternary diagram form a tetrahedron of height $A$. The four vertices of the diagram represent $w=A$, $x=A$, $y=A$, and $z=A$, while the opposing faces are surfaces on which $w=0$, $x=0$, $y=0$, and $z=0$, respectively. At each point inside the tetrahedron, the value of $w$ is given by perpendicular distance to the $w=0$ face, and the values of the other components are defined analogously. Equilateral tetrahedrons have the property that the sum of the distances from any interior point to each of the four faces equals the height of the tetrahedron $A$. We are thus assured that $w+x+y+z=A$ is satisfied at every point within the quaternary diagram.
		
	We use quaternary diagrams to plot all the possible ways a planet of mass $M_p$ and radius $R_p$ can be partitioned into the four layers of our fiducial model described in \S~\ref{sec:model}. In this case, the four-component data that we are plotting in the diagram are the fractions of the mass of the planet in each of the four interior layers $\left(x_{core}, x_{mantle}, x_{\rm{H_2O}}, x_{\rm{H/He}}\right)$, which are constrained to sum to unity. The summits of the tetrahedron represent extreme cases in which the planet is 100\% iron, 100\% silicates, 100\% water ices, or 100\% H/He. The face opposite the H/He summit turns out to be a ternary diagram for the gas-less interior compositions of the planet.

\section{Results}
\label{sec:results}

	Our eventual aim is to draw robust conclusions about the composition of a low-mass exoplanet by fully exploring and quantifying the associated uncertainties. There is an inherent degeneracy in the planetary compositions that can be inferred from planet mass and radius measurements alone; for a specified planet mass, many different distributions of matter within the planet interior layers  can produce identical radii. In planet interior models incorporating $N$ distinct chemical layers, specifying a planet mass and radius each impose a constraint on the layer masses, leaving $\left(N-2\right)$ degrees of freedom in the allowed compositions $\{x_i\}$. Further compositional uncertainties may be introduced if the planetary energy budget or chemical makeup are not well known and if significant measurement uncertainties are present in observationally derived parameters. 	

In this work, we examine the constraints that can be placed on a transiting exoplanet's interior using only structural models for the planet. By not employing planet formation arguments to impose further constrain the planetary compositions,  our results remain largely independent of planet formation theories. In this section, we apply our interior structure model to examine the possible compositions of several example planets: CoRoT-7b, GJ~581d, GJ~436b, and HAT-P-11b.

\subsection{CoRoT-7b}
\label{sec:CoRoT7b}

The recent discovery of the first transiting super-Earth, CoRoT-7b, has ushered in a new era of exoplanet science \citep{LegerEt2009A&A}. CoRoT-7b is on a $0.85359\pm0.00005$ day period around a bright $V=11.7$ G9V star. The host star is very active which complicates measurement of the transiting planet's mass and radius. By forcing the stellar radius to be $R_*=(0.87\pm0.04)~R_{\odot}$, a planetary radius of $R_p=(1.68\pm0.09)~R_{\oplus}$ is derived from the transit depth \citep{LegerEt2009A&A}. The planetary nature of CoRoT-7b has recently been confirmed by Doppler measurements revealing a planetary mass of $M_p=(4.8\pm0.8)~M_{\oplus}$ \citep{QuelozEt2009A&A}. For the very first time, both the mass and radius of a super-Earth sized exoplanet have been measured, thereby offering the first hints about the interior composition of a planet in the mass range between Earth and Neptune.

In this section we do not consider the possibility that CoRoT-7b could harbor a gas layer or a significant water ocean. With an orbital semimajor axis of  $a=(0.0172\pm0.00029)~\rm{AU}$  (about four stellar radii), CoRoT-7b is receiving an extreme amount of stellar irradiation. CoRoT-7b is most likely tidally locked,  with a temperature of up to $2560\pm125~\rm{K}$ at the sub-stellar point assuming an albedo of $A=0$ and no energy redistribution \citep{LegerEt2009A&A}. Limits on the lifetime of a gas layer or a water ocean under such extreme radiation are discussed in \S~\ref{sec:Dis}. We focus here on what we can learn about the composition of CoRoT-7b if it is a  purely dry, gas-less telluric planet. \citet{ValenciaEt2009AstroPh} offer another point of view, considering the possibility of an H/He or vapor atmosphere on CoRoT-7b.

\begin{figure}
\epsscale{1.15}
\plotone{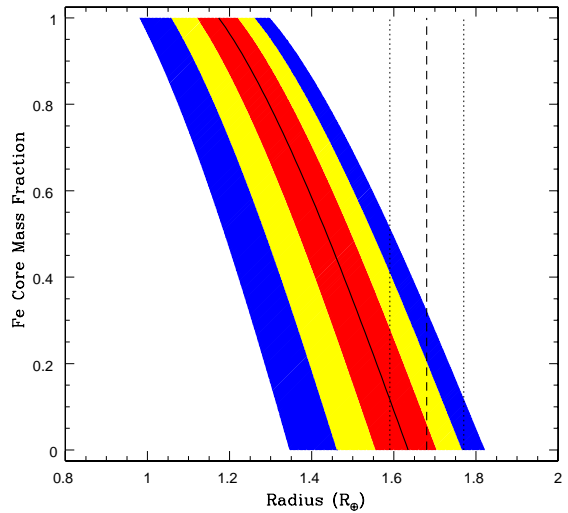}
\caption{CoRoT-7b core mass fraction as a function of planetary radius. The planetary mass is $(4.8\pm0.8)~M_{\oplus}$. We neglect the possible presence of water or a gas layer and consider a two-layer planet comprised of a pure iron core surrounded by a  $\rm{Mg_{0.9}Fe_{0.1}SiO_3}$ mantle. The red, yellow and blue shaded regions denote the core mass fractions obtained when varying the CoRoT-7b mass within its $1~\sigma$, $2~\sigma$, and $3~\sigma$ error bars, respectively.  The black vertical lines delimit the measured radius $R=(1.68\pm0.09)~R_{\oplus}$ (dashed) and its $1~\sigma$ error bar (dotted).}
\label{fig:CoRoT7bcmf}
\end{figure}

We first examine the interior composition of COROT-7b under the assumption of an iron core and a mantle composed of silicate perovskite ($\rm{Mg_{0.9}Fe_{0.1}SiO_3}$,  approximately similar to Earth's mantle). When considering only two compositional layers, the measured mass and radius uniquely determine the two layer masses. The core mass fraction as a function of planet radius for CoRoT-7b is displayed in Figure~\ref{fig:CoRoT7bcmf}. The solid black line denotes the fraction of CoRoT-7b's mass in its iron core assuming the fiducial planetary mass $M_p=4.8~M_{\oplus}$, while the red, yellow and blue shaded regions delimit the $1~\sigma$, $2~\sigma$, and $3~\sigma$ error bars on $M_p$ respectively. The measured planet radius and its $1~\sigma$ error bars are denoted by the dashed and dotted black vertical lines, respectively. An Earth-like composition, having $30\%$ of its mass in an iron core and the remaining $70\%$ of its mass in a silicate mantle, is consistent with the measured mass and radius for CoRoT-7b within $1~\sigma$. If the CoRoT-7b core is not pure iron but also contains a light element, the core mass fraction at a specified planetary radius will be larger. Including 10\% sulfur by mass in the iron core EOS increases the CoRoT-7b radius by $0.08~R_{\oplus}$ at a core mass fraction of 1 (at the top of Figure~\ref{fig:CoRoT7bcmf}), while having no effect on the radius at a core mass fraction of 0 (at the bottom of Figure~\ref{fig:CoRoT7bcmf}).  

Our interior structure model can strengthen the observational constraints on CoRoT-7b's mass and radius. With the assumption that CoRoT-7b does not have a significant water or gas layer, some of the mass-radius pairs within $M_p\pm1\sigma_M$ and $R_p\pm1\sigma_R$ (including the fiducial $0~\sigma$ mass-radius pair) can be ruled out because they correspond to bulk densities lower than a pure silicate planet. These excluded mass-radius pairs would necessitate water (or some other component lighter than perovskite) in their composition. The fact that some $1\sigma$ CoRoT-7b mass-radius pairs are excluded can be seen from Figure~\ref{fig:CoRoT7bcmf},  where the red band denoting planetary masses within $1~\sigma$ of the measured value never fully crosses the $R_p+1\sigma_R$ dotted line even at a 100\% perovskite composition. While most of this work is devoted to constraining a planet's interior structure from mass and radius measurements, this is an example of how limits on a planet's interior structure could be used to improve our constraints on a planet's mass and radius. 

\begin{figure}
\epsscale{1.15}
\plotone{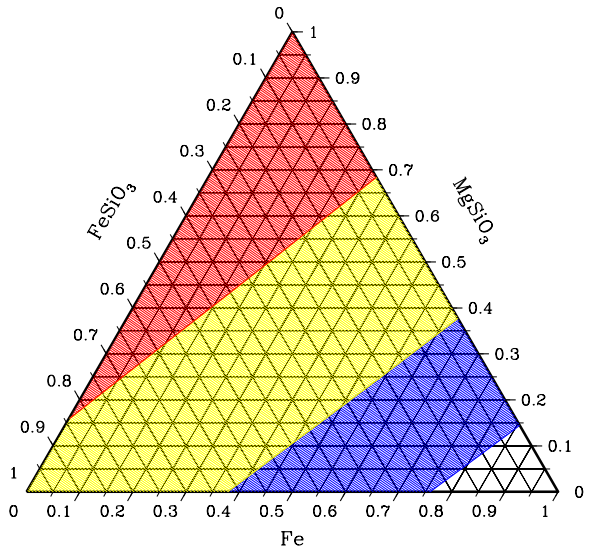}
\caption{CoRoT-7b ternary diagram. Plausible compositions for CoRoT-7b are shown, provided the planet has no interior water and no H/He layer. The fractions of the planet's mass in the Fe core, in MgSiO$_3$, and in FeSiO$_3$ are plotted on the three axes. MgSiO$_3$ and FeSiO$_3$ are mixed together in the mantle as $\rm{Mg}_{1-\chi}\rm{Fe}_{\chi}$SiO$_3$.}
\label{fig:CoRoTnoH2OTern}
\end{figure}

The amount of iron in a exoplanetary mantle is not known. Earth's mantle has about 10\% iron and 90\% Mg by number fraction ($\rm{Mg_{0.9}Fe_{0.1}SiO_3}$), but exoplanets may have varying amounts. \citet{ElkinsTanton&Seager2008bApJ} describe an extreme example of a coreless terrestrial planet in which all of the planet's iron is mixed in the mantle instead of sequestered in the core. To explore the effect of varying the mantle iron fraction we present a ternary diagram in Figure~\ref{fig:CoRoTnoH2OTern} that shows the tradeoff between the mass of iron in the mantle compared to the mass of iron in the core. The fractions of the planet's mass in the Fe core, in MgSiO$_3$ and in FeSiO$_3$ are plotted on the three axes. MgSiO$_3$ and FeSiO$_3$ are mixed together in the mantle as Mg$_{1-\chi}$Fe$_\chi$SiO$_3$, where $\chi$ is the number fraction of FeSiO$_3$. The red, yellow and blue shaded regions denote interior compositions that are consistent with the measured planetary mass and radius to within $1~\sigma$, $2~\sigma$, and $3~\sigma$ of the observational uncertainties respectively. All the ternary diagram except the high Fe corner ($x_{\rm{Fe}}\gtrsim0.76-0.86$) is shaded to within $3~\sigma$. Because FeSiO$_3$ and MgSiO$_3$ have similar densities (compared to the density contrast between pure Fe and perovskite), we have very little ability to discriminate the iron content of the mantle from a mass and radius measurement alone. Nonetheless, $\chi$ contributes to the uncertainty in the core mass fraction of CoRoT-7b. 

\subsection{GJ~581d}
\label{sec:GJ581d}

We now consider the super-Earth exoplanets that are large and cool enough that they might retain a small hydrogen-helium gas layer. As an example we use GJ~581d, a $M_p\sin{i}=7.09~M_{\oplus}$ super-Earth with a semimajor axis $a=0.22~\rm{AU}$ that is part of a multi-planet system around an $L=0.013~L_{\odot}$ M3 dwarf star~\citep{UdryEt2007A&A, MayorEt2009aA&A}. GJ~581 is estimated to be $8^{+3}_{-1}~\rm{Gyr}$ old\footnote{\url{exoplanet.eu}}. The radius of GJ~581d has not yet been measured. In this section, we adopt the minimum mass for GJ~581d and consider two different possible planetary radii: $R_p=1.5$ and $2.0~R_{\oplus}$. While these values may not represent the properties of the true GJ~581d planet, we use them to illustrate how the possible presence of a gas layer and observational uncertainties will affect our ability to make inferences about the interior composition of transiting super-Earths. 

\begin{figure}
\epsscale{1.15}
\plotone{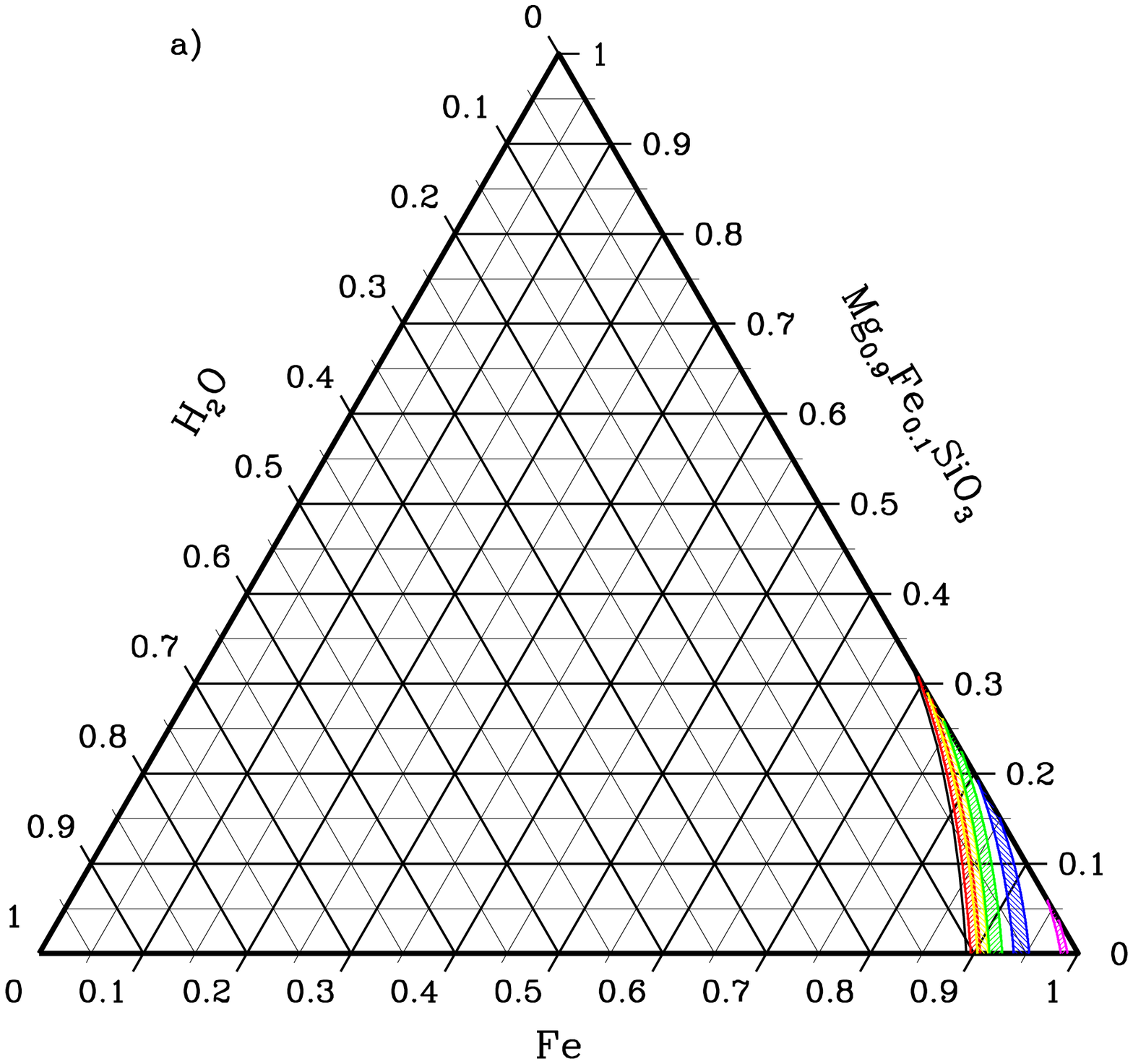}
\plotone{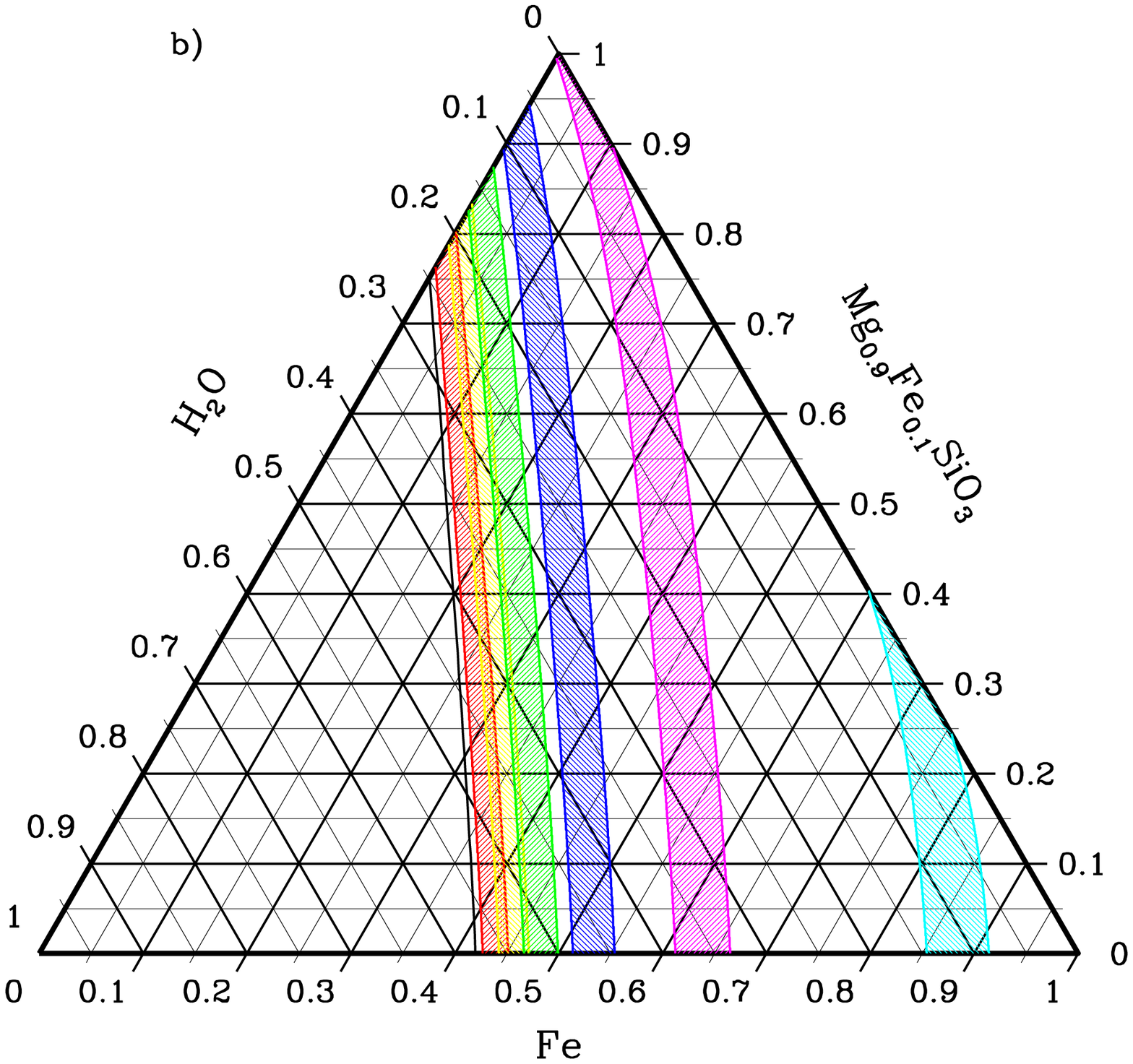}
\caption{Ternary diagram for the solid core of GJ~581d. The GJ~581d minimum mass  $M_p=7.09~M_{\oplus}$ is assumed. Each diagram represents a different possible planetary radius: (a) $R_p=1.5 R_{\oplus}$, and (b) $R_p=2.0 R_{\oplus}$. The relative contributions of the iron core,  $\rm{Mg_{0.9}Fe_{0.1}SiO_3}$ mantle, and $\rm{H_2O}$ ices to the mass of the solid planet bulk are plotted. The leftmost black curve represents the locus of gas-less compositions, and gas mass fraction increases to the right toward the Fe vertex. The different colored bands designate various gas mass fractions $\left(x_{H/He}\right)$: $10^{-7}$ (red), $10^{-6}$  (yellow), $10^{-5}$ (green), $10^{-4}$ (blue), $10^{-3}$ (magenta), and $10^{-2}$ (cyan). For reference, the Earth's gas mass fraction is about $10^{-6}$ and Venus' is about $10^{-4}$. The width of each of the colored bands is produced by varying the atmospheric parameters within the ranges $\gamma=0.1-10$, $T_{0}=181-285$, and $T_{\rm{eff}}=73-93$ ($R_p=1.5 R_{\oplus}$) or $T_{\rm{eff}}=59-75$ ($R_p=2.0 R_{\oplus}$) .}
\label{fig:Gl581dgasTern}
\end{figure}

The two putative planetary radii considered for GJ~581d lead to interior compositions having very different characteristics. Ternary diagrams assuming a radius of $R_p=1.5$ and $2.0~R_{\oplus}$ for GJ~581d are displayed in Figures~\ref{fig:Gl581dgasTern}(a) and (b), respectively. The leftmost black curve in each diagram represents the locus of possible gas-less compositions for the stipulated mass and radius. The $R_p=1.5~R_{\oplus}$ planet is very dense and iron-rich; it could have a Mercury-like composition with $68\%$ of its mass in an iron core enveloped by a silicate mantle accounting for the remaining $32\%$ of the mass.  In contrast, possible gas-less compositions for $R_p=2.0~R_{\oplus}$ are all icy planets with 25\%-58\% $\rm{H_2O}$ by mass.

In Figure~\ref{fig:Gl581dgasTern}, each colored band designates a different gas mass fraction. For non-zero gas mass fractions $\left(x_{gas}\right)$, the relative contributions of the iron core,  $\rm{Mg_{0.9}Fe_{0.1}SiO_3}$ mantle, and $\rm{H_2O}$ ices to the solid interior (inner three layers) of GJ~581d are plotted; effectively, the fraction of the planet mass in each of the solid layers is re-normalized by $\left(1-x_{\rm{H/He}}\right)$. The non-zero width of the gas mass fraction bands in the ternary diagrams is due to the uncertainty in the atmospheric \textit{P-T} profile. Following the scheme described in \S~\ref{sec:Ebudget}, we consider $\gamma=0.1-10$, $T_{0}=181-285$, $T_{\rm{eff}}=73-93$  for $R_p=1.5 R_{\oplus}$, and $T_{\rm{eff}}=59-75$ for $R_p=2.0 R_{\oplus}$. 

Allowing for the presence of a gas layer on GJ~581d significantly increases the range of interior compositions that can produce the stipulated mass and radius. The more gas GJ~581d contains, the higher the average density of the inner three layers must be to still satisfy the planetary mass and radius constraints. More gas results in an increase in the proportion of iron, as manifested in the ternary diagram (Figure~\ref{fig:Gl581dgasTern}) by the fact that the gas mass fraction increases to the right toward the Fe vertex. An upper limit on the mass of gas that GJ~581d can support is reached if the planet has no $\rm{H_2O}$ or silicates but consists solely of H/He enveloping an iron core (a composition corresponding to the iron vertex, Figure~\ref{fig:Gl581dgasTern}). This H/He mass upper limit occurs at $0.12\%-0.19\%$ for  $R_p=1.5~R_{\oplus}$ and at $1.7\%-2.2\%$ for $R_p=2.0~R_{\oplus}$. These limits consider only the constraints imposed by the planetary mass and radius and not the lifetime of the gas layer to atmospheric escape. Having a gas layer contribute $10^{-5}$ of the mass of GJ~581d (for comparison the Earth's atmosphere is about $10^{-6}$ of an Earth mass) increases the minimum iron core mass fraction for a $R_p=1.5~R_{\oplus}$ planet from the $68\%$ gas-less value to $74\%-78\%$ and decreases the minimum $\rm{H_2O}$ mass fraction for a $R_p=2.0~R_{\oplus}$ planet from the 25\% gas-less value to $13\%-17\%$. Although a gas layer on GJ~581d can make at most a small contribution to the planetary mass, it can nonetheless have a very important effect on the allowed proportions of the inner three layers and on our ability to infer the planet's interior composition. 

\begin{figure}
\epsscale{1.15}
\plotone{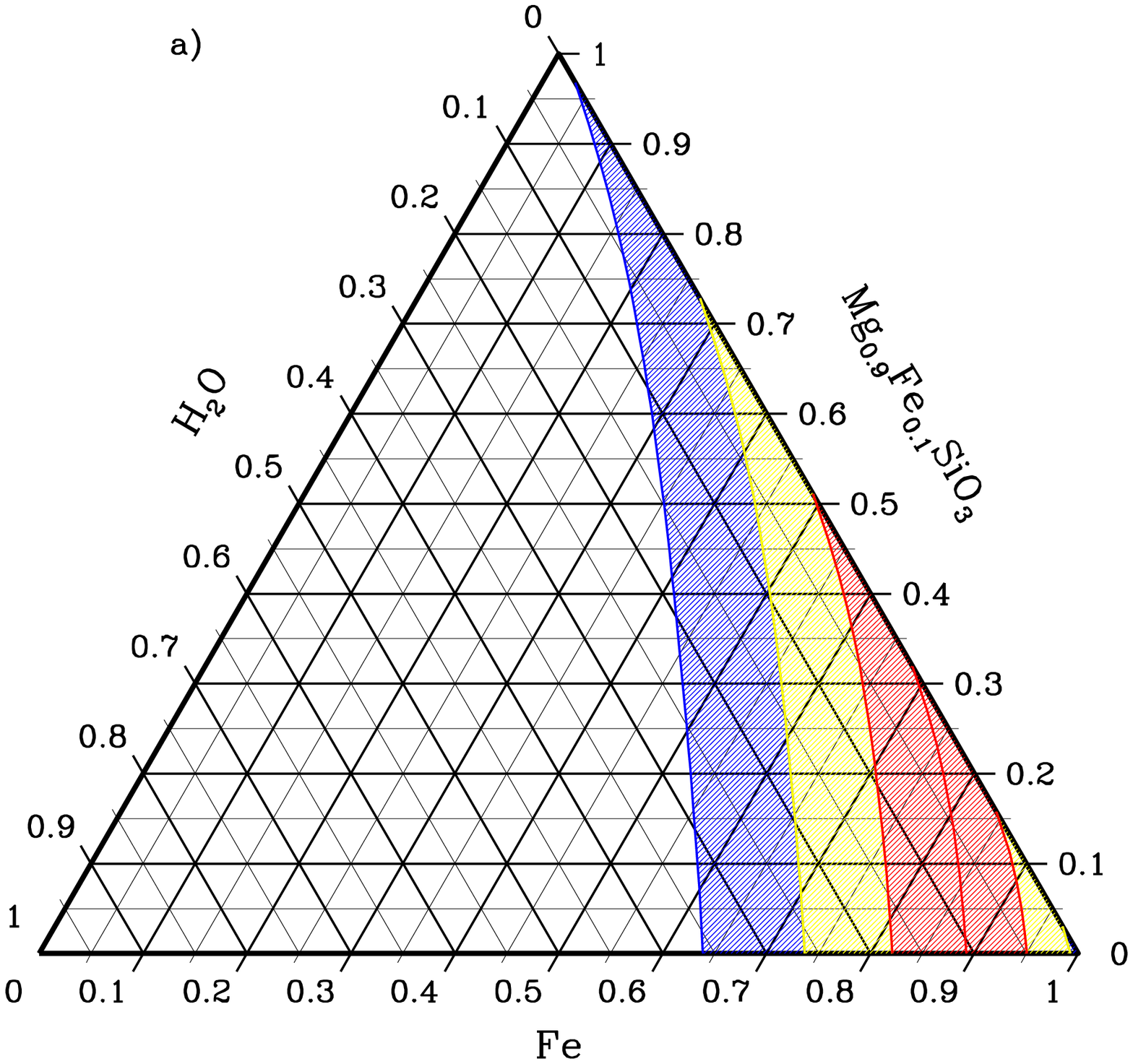}
\plotone{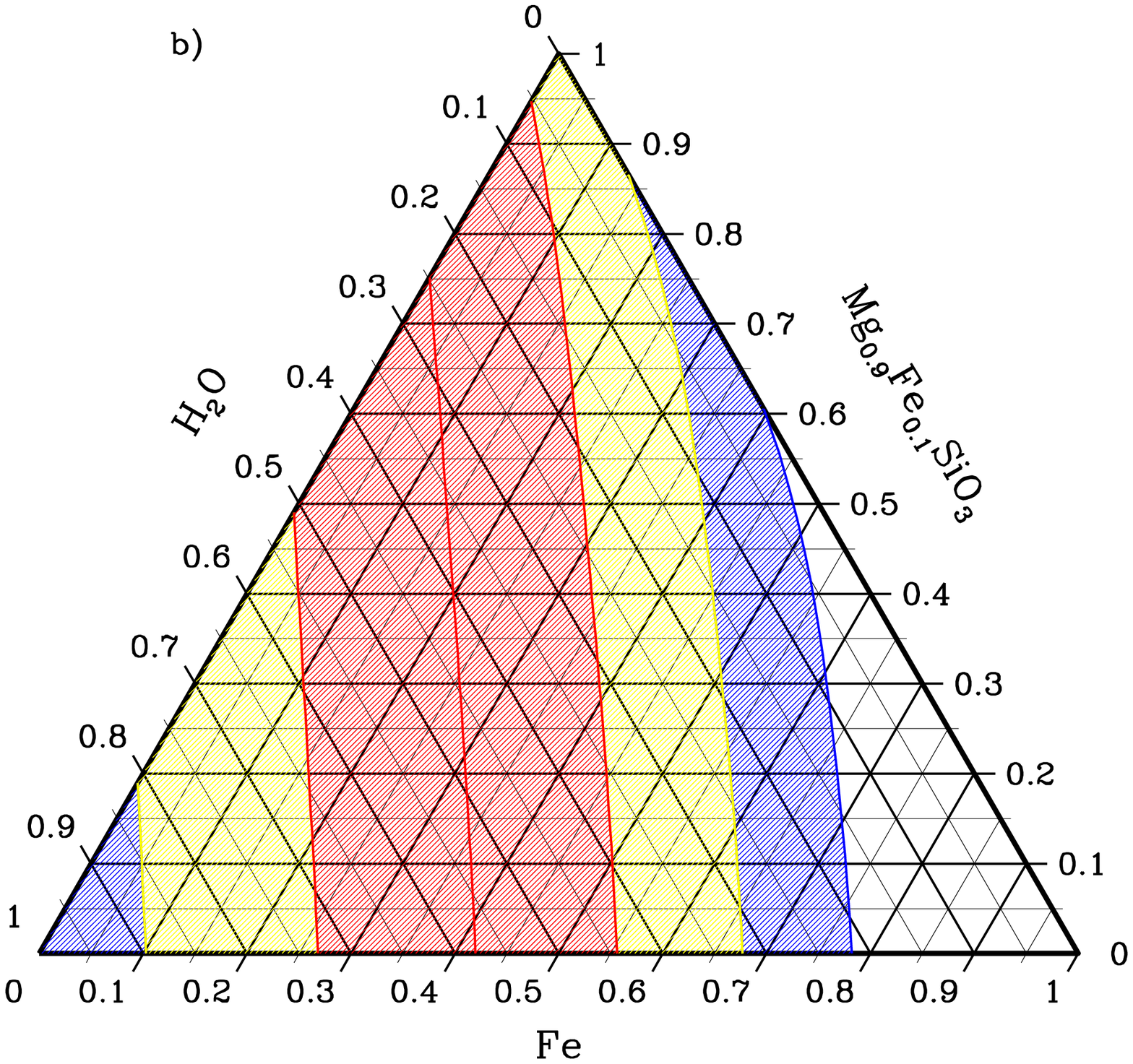}
\caption{Ternary diagram displaying plausible gas-less compositions for GJ~581d. An observational uncertainty of 5\% is included on both the assumed mass  $\left(M_p=7.09~M_{\oplus}\right)$  and the assumed radii. Each diagram represents a different possible planetary radius: (a) $R_p=1.5 R_{\oplus}$, and (b) $R_p=2.0 R_{\oplus}$. The red, yellow, and blue shaded regions denote compositions that are consistent with $M_p$ and $R_p$ to within $1~\sigma$, $2~\sigma$, and $3~\sigma$, respectively.}
\label{fig:Gl581dTern5}
\end{figure}

So far we have only considered the inherent uncertainty in the composition of GJ~581d that could be inferred from a planetary mass and radius. In practice, observational uncertainties also impact our ability to constrain the interior composition of a transiting super-Earth. For illustration purposes, we consider the same two putative GJ~581d mass-radius pairs, and assume an optimistic but plausible uncertainty of 5\% on both the planetary mass and radius. Ternary diagrams plotting gas-less compositions consistent with the planet mass and radius to within  $1~\sigma$, $2~\sigma$, and $3~\sigma$ are shown in Figure~\ref{fig:Gl581dTern5}(a) for $R_p=1.5 R_{\oplus}$ and Figure~\ref{fig:Gl581dTern5}(b)  for $R_p=2.0 R_{\oplus}$. If compositions including gas layers were included in Figure~\ref{fig:Gl581dTern5}, the shaded $n~\sigma$ regions would all be smeared out to the right and extended to the Fe vertex. 

Even neglecting the effect of a possible gas layer, the interior composition of GJ~581d is far better constrained for a radius of $R_p=\left(1.5\pm5\%\right) R_{\oplus}$ (Figure~\ref{fig:Gl581dTern5}(a)) than it is for $R_p=\left(2.0\pm5\%\right) R_{\oplus}$ (Figure~\ref{fig:Gl581dTern5}(b)).  The superior compositional constraints attained at the smaller planetary radius are a consequence of two effects. First, the $R_p=1.5 R_{\oplus}$ planet has a lower inherent compositional degeneracy for its fiducial ($0~\sigma$) mass and radius. The $R_p=1.5~R_{\oplus}$  planet is dense enough that it must contain a large amount of iron, while the $R_p=2.0~R_{\oplus}$  has a more intermediate density and could be assembled from a wider range of combinations of iron, silicates, and water. This can be seen from the ternary diagrams (Figure~\ref{fig:Gl581dTern5}) in which the line representing the gas-less compositions for $\left(M_p=7.09~M_{\oplus},\; R_p=1.5 R_{\oplus}\right)$ is much shorter than the line representing the possible gas-less compositions for $\left(M_p=7.09~M_{\oplus},\; R_p=2.0 R_{\oplus}\right)$. Second, the separation in the $1~\sigma$, $2~\sigma$, and $3~\sigma$ contours of the ternary diagram are much wider in the case of $R_p=2.0 R_{\oplus}$  in Figure~\ref{fig:Gl581dTern5}(b) than they are for $R_p=1.5 R_{\oplus}$ in Figure~\ref{fig:Gl581dTern5}(a). The relative uncertainty on the average planet density $\bar\rho$ is identical (to first order) for both GJ~581d radii considered $\left(\Delta\bar\rho/\bar\rho \approx \sqrt{\left(\Delta M/M\right)^2+\left(3\Delta R/R\right)^2}=16\%\right)$, while the spacings between iso-mass and radius curves on the ternary diagram are roughly proportional to $\propto\Delta\bar\rho/\bar\rho^2$ \citep{Zeng&Seager2008PASP}. Thus, the separation in the $1~\sigma$, $2~\sigma$, and $3~\sigma$ contours of the ternary diagram increases with decreasing planetary density. This example illustrates how our ability to constrain the interior composition of a transiting super-Earth depends not only on the precision of our measurements, but also on the true mass and radius of the planet. For a given relative uncertainty on the average planet density, the composition can be best constrained for very dense planets (near the Fe vertex).

\subsection{GJ~436b}

GJ~436b, a hot Neptune orbiting a nearby M star \citep{ButlerEt2004ApJ, ManessEt2007PASP}, was the first known transiting intermediate-mass planet. Since GJ~436b was found to transit its star by \citet{GillonEt2007A&Aa}, substantial efforts have been made to measure its mass and radius using  photometric data from \textit{the Spitzer Space Telescope} \citep{DemingEt2007ApJ, GillonEt2007A&Ab}, from  \textit{the Hubble Space Telescope} \citep{BeanEt2008A&A}, and from further ground-based observations \citep[e.g.][]{ShporerEt2009ApJ}. Here, we adopt values for the properties of GJ~436b and its host star given by \citet{Torres2007ApJ} and \citet{TorresEt2008ApJ}, who employed a weighted average of light-curve parameters from ground-based \citep{GillonEt2007A&Aa} and \textit{Spitzer} studies \citep{DemingEt2007ApJ, GillonEt2007A&Ab}:  $L_*= 0.0260^{+0.0014}_{-0.0017}~L_{\odot}$, $M_p=23.17\pm0.79 M_{\oplus}$, $R_p=4.22^{+0.09}_{-0.10} R_{\oplus}$, and $a=0.02872^{+0.00029}_{-0.00026}~\rm{AU}$.

\begin{figure}
\epsscale{1.15}
\plotone{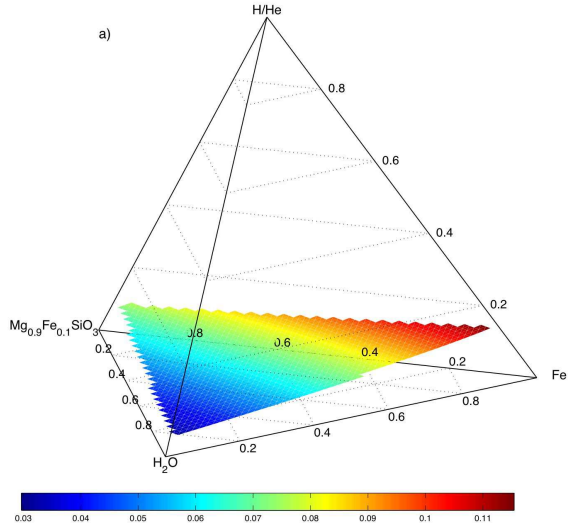}
\plotone{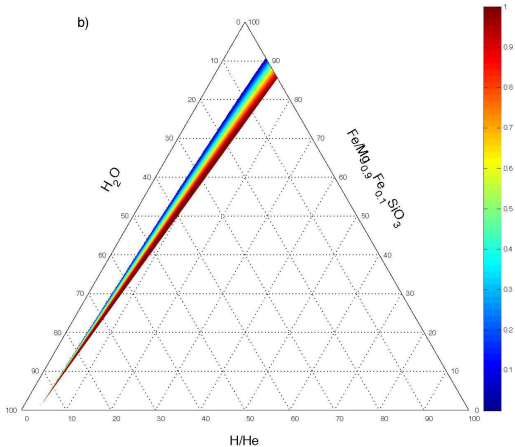}
\caption{Fiducial GJ~436b quaternary and ternary diagrams. The allowed compositions of GJ~436b for our fiducial choice of structural and atmospheric parameters ($M_p=23.17 M_{\oplus}$, $R_p=4.22 R_{\oplus}$, $T_0=663~\rm{K}$, $T_{\rm{eff}}=70~\rm{K}$, $\gamma=1$) are shown. In panel (a) we show a three-dimensional quaternary diagram plotting the fraction of the planet's mass in the iron core, $\rm{Mg_{0.9}Fe_{0.1}SiO_3}$ mantle, water ices, and H/He gas layer. The surface is colored according to the fraction of the mass of the planet found in the gas layer. Panel (b)  displays the same data as (a)  in a two-dimensional ternary diagram. In panel (b)  the core and mantle are combined together on a single axis, with the vertical distance from the upper vertex determined by the fraction of the planet's mass in the two innermost planet layers. The color shading denotes the relative contribution of the core to the total mass in the inner two layers. The width of the shaded wedge of allowed compositions is due to varying the ratio of Fe to $\rm{Mg_{0.9}Fe_{0.1}SiO_3}$: the blue edge of the allowed compositions represents planets having no Fe, while the red edge represents planets lacking $\rm{Mg_{0.9}Fe_{0.1}SiO_3}$.}
\label{fig:GJ436bFid}
\end{figure}

\begin{figure}
\epsscale{1.15}
\plotone{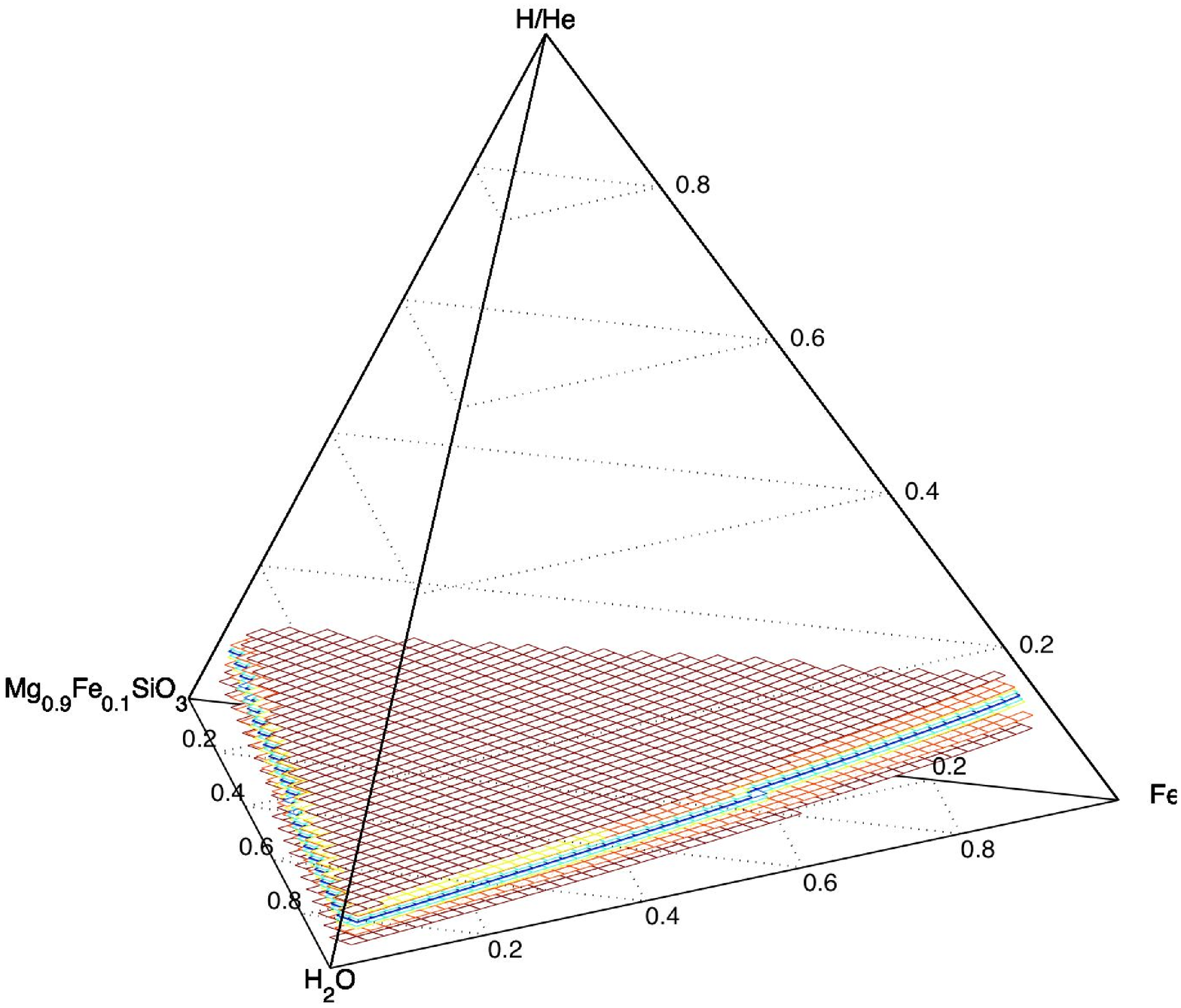}
\plotone{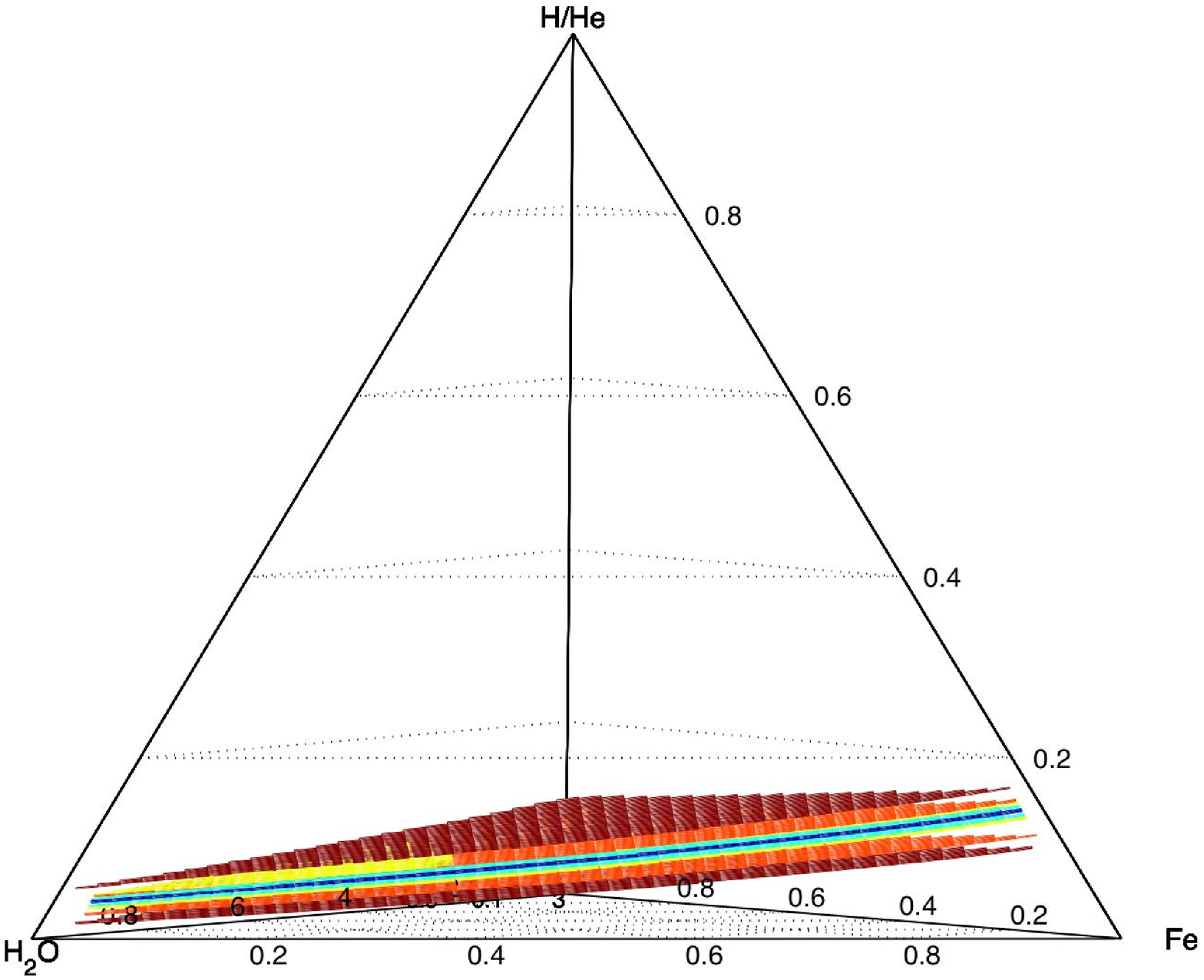}
\caption{GJ~436b quaternary diagram. Both model and observational uncertainties are taken into account to determine the plausible interior compositions of GJ~436b in this diagram. Two different views of the same quaternary diagram are shown. The surface of allowed compositions for our fiducial choice of model parameters ($M_p=23.17 M_{\oplus}$, $R_p=4.22 R_{\oplus}$, $T_0=663~\rm{K}$, $T_{\rm{eff}}=70~\rm{K}$, $\gamma=1$) is displayed in navy blue; this surface is the same as displayed in the quaternary diagram in Figure~\ref{fig:GJ436bFid}(a). To explore how uncertainties in model parameters weaken the constraints that can be placed on GJ~436b's interior composition, we vary each model parameter in turn while keeping all others fixed at their fiducial values. Two surfaces of the same color delimit the volume of composition space that is consistent with the range of values examined for each parameter. We consider $\gamma=0.1-10$ (cyan),  $T_{0}=937-595~\rm{K}$ (green), and $T_{\rm{eff}}=58-113~\rm{K}$ (orange). The yellow surfaces denote the effect of varying the planet mass and radius within their $1\sigma$ observational uncertainties while maintaining all other model parameters at their fiducial values. Finally, the red surfaces delimit the full volume of possible compositions obtained by varying all parameters within the ranges described above.}
\label{fig:GJ436bQuatParam}
\end{figure}

The measured mass and radius of GJ~436b constrain its bulk interior composition. Allowed compositions for our fiducial planetary parameters ($M_p=23.17 M_{\oplus}$, $R_p=4.22 R_{\oplus}$, $T_0=663~\rm{K}$, $T_{\rm{eff}}=70~\rm{K}$, $\gamma=1$) are displayed in Figure~\ref{fig:GJ436bFid}.  For our fiducial set of GJ~436b model parameters, the allowed compositions form a two-dimensional surface in the quaternary diagram (Figure~\ref{fig:GJ436bFid}(a)). This illustrates the inherent compositional degeneracy originating from an underconstrained interior model; the measured mass and radius place only two constraints on the masses in each of the four interior layers. When uncertainties in the model parameters are considered, the surface of allowed compositions gains some thickness and spreads into a volume, weakening the constraints that can be placed on GJ~436b's composition (Figure~\ref{fig:GJ436bQuatParam}). Not all of the quaternary diagram is filled, however, even when both observational and model uncertainties are taken into account. Some interior compositions (specifically those outside the red surfaces in Figure~\ref{fig:GJ436bQuatParam}) can thus be ruled out for GJ~436b.

GJ~436b can support a range of gas mass fractions, but must have some gas. For our fiducial parameter choices, GJ~436b could be between 3.6\% and 14.5\% gas by mass. The gas mass fraction needed to produce the observed transit depth depends on the composition of the planet's solid core: water worlds with large ice layers fall near the minimum gas mass fraction (3.6\%), while dry planets with iron-rich cores require up to 14.5\% gas. The tradeoff between H/He and water contents is illustrated in Figure~\ref{fig:GJ436bFid}(b), in which the iron core and perovskite mantle are combined together on one axis to form a ternary diagram from the data presented in Figure~\ref{fig:GJ436bFid}(a). In Figure~\ref{fig:GJ436bFid}(b), the shaded wedge of allowed compositions slopes from near the pure $\rm{H_2O}$ vertex toward increasing H/He and the opposite 0\% water edge. Because the allowed compositions span almost the entire $\rm{H_2O}$ axis (from 0\% to 96.4\%), the mass fraction of water on GJ~436b is poorly constrained by the measured mass and radius alone.  

\begin{figure}
\epsscale{1.15}
\plotone{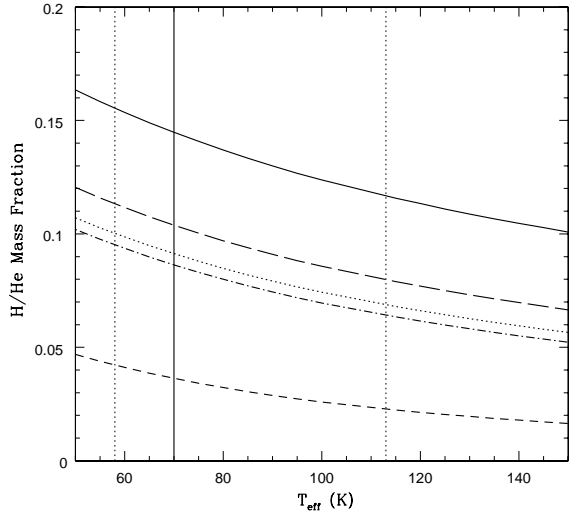}
\caption{Gas mass fraction of GJ~436b as a function of $T_{\rm{eff}}$. All parameters other than $T_{\rm{eff}}$ (including $M_p$, $R_p$, $T_0$, and $\gamma$) are fixed at their fiducial values. Curves for different end member compositions of the solid bulk of GJ~436b below the H/He layer are displayed: pure iron (solid), pure perovskite $\rm{Mg_{0.9}Fe_{0.1}SiO_3}$ (dotted), pure water (short dashed), 25\% iron 75\% perovskite (long dashed), and 25\% iron 50\% perovskite 25\% water (dot-dashed). The solid vertical line denotes the fiducial value of $T_{\rm{eff}}=70~\rm{K}$, while the vertical dotted lines delimit the range of $T_{\rm{eff}}$ values considered (58-1130~K). }
\label{fig:GJ436bgmfTeff}
\end{figure}

The range of gas mass fractions that can be supported by GJ~436b strongly depends on the internal heat flux as parameterized by $T_{\rm{eff}}$. At higher temperatures, the gas layer is less dense and both the minimum and maximum gas mass fractions decrease, while at lower temperatures the gas layer is more dense and the gas mass fraction extremes both increase. For instance, at $T_{\rm{eff}}=113~\rm{K}$ allowed gas mass fractions range from 2.3\% to 11.7\%, while at $T_{\rm{eff}}=58~\rm{K}$ GJ~436b must be between 4.2\% and 15.5\% gas by mass. Figure~\ref{fig:GJ436bgmfTeff} plots the gas mass fraction of GJ~436b as a function of $T_{\rm{eff}}$ for various interior compositions (with all parameters other than $T_{\rm{eff}}$ fixed at their fiducial values). Using the formalism described in \S\ref{sec:Ebudget}, we estimate $T_{\rm{eff}}=70^{+43}_{-12}~\rm{K}$ for a planet age of $6^{+4}_{-5}~\rm{Gyr}$; the age of the GJ~436 solar system is essentially unconstrained by observations since GJ~436 is unevolved on the main sequence \citep{Torres2007ApJ}.  Any constraints placed on the interior composition of GJ~436b will be sensitive to assumptions made about the intrinsic luminosity of the planet.  

Out of all the atmospheric parameters in our model, uncertainties in $T_{\rm{eff}}$ have the most important effect on limiting the compositional constraints that can be placed on GJ~436b. In Figure~\ref{fig:GJ436bQuatParam}, we explore the effect each model parameter has on the volume of allowed compositions while keeping all other parameters fixed at their fiducial values. Varying $T_{\rm{eff}}$ from 58 to 113~K expands the space of allowed GJ~436b compositions far more than varying $T_{0}=595-937~\rm{K}$ or $\gamma=0.1-10$. The relative importance of the $T_{\rm{eff}}$ parameter was not unexpected. The intrinsic luminosity determines the asymptotic behavior of the \citet{Hansen2008ApJS} temperature profile in the radiative regime at larger optical depths ($\tau\gtrsim\left(T_0/T_{\rm{eff}}\right)^4$). While $\gamma$ and $T_0$ affect the temperature profile in the outer low-density low-optical-depth region of the gas layer, the intrinsic luminosity $T_{\rm{eff}}$ dominates in the higher density inner regions of the radiative gas layer. As a result, $T_{\rm{eff}}$ affects a larger component of the gas layer mass and exerts a larger influence on the transition to a convective gas layer and the entropy of the interior adiabat.  \citet{AdamsEt2008ApJ} also used the temperature profile from \citet{Hansen2008ApJS} and similarly found that $T_{\rm{eff}}$ had the largest effect on their simulated planet radii. 

Observational uncertainties dominate most of the model uncertainties discussed above. The $1\sigma$ observational uncertainties on mass and radius are second only to the uncertainty in the planetary internal heat flux $T_{\rm{eff}}$ in their effect on our ability to constrain the interior composition of GJ~436b. This is evident from Figure~\ref{fig:GJ436bQuatParam} by comparing the yellow surfaces delimiting the volume of compositions obtained by varying the GJ~436b mass and radius within their $1\sigma$ error bars and the orange surfaces denoting the effect of uncertainties in $T_{\rm{eff}}$. In this case, he range of plausible $T_{\rm{eff}}$ would have to be constrained to better than about 20\% of its fiducial value before the observational uncertainties in the planet radius would dominate the thickness of the volume of allowed compositions. More theoretical work is required to model the cooling and internal heat flux of hot Neptunes and super-Earths harboring significant gas layers. Until progress is made in constraining $T_{\rm{eff}}$, improvements in the observational uncertainties on the GJ~436b mass and radius will not translate into substantial improvements in our ability to constrain the GJ~436b interior composition. 

\subsection{HAT-P-11b}
\label{sec:HATP11b}

HAT-P-11b is the first hot Neptune to be discovered by transit searches~\citep{BakosEt2010ApJ}. HAT-P-11b existence has since been confirmed by \citet{DittmannEt2009ApJ}. Orbiting at $a=0.0530^{+0.0002}_{-0.0008}~\rm{AU}$ from a K4 dwarf start with ${T_{\rm{eff}}}_{*}=4780\pm50~\rm{K}$, HAT-P-11b is similar to GJ~436b in mass and radius: $M_p=25.8\pm2.9 M_{\oplus}$ and $R_p=4.73\pm0.16 R_{\oplus}$ \citep{BakosEt2010ApJ}. Its host star is HAT-P-11 is $\rm{6.5^{+5.9}_{-4.1}~Gyr}$  old \citep{BakosEt2010ApJ}, as determined from Yale-Yonsei isochrones \citep{YiEt2001ApJS}. To date, HAT-P-11b and GJ~436b are the only known transiting hot Neptunes.

\begin{figure}
\epsscale{1.15}
\plotone{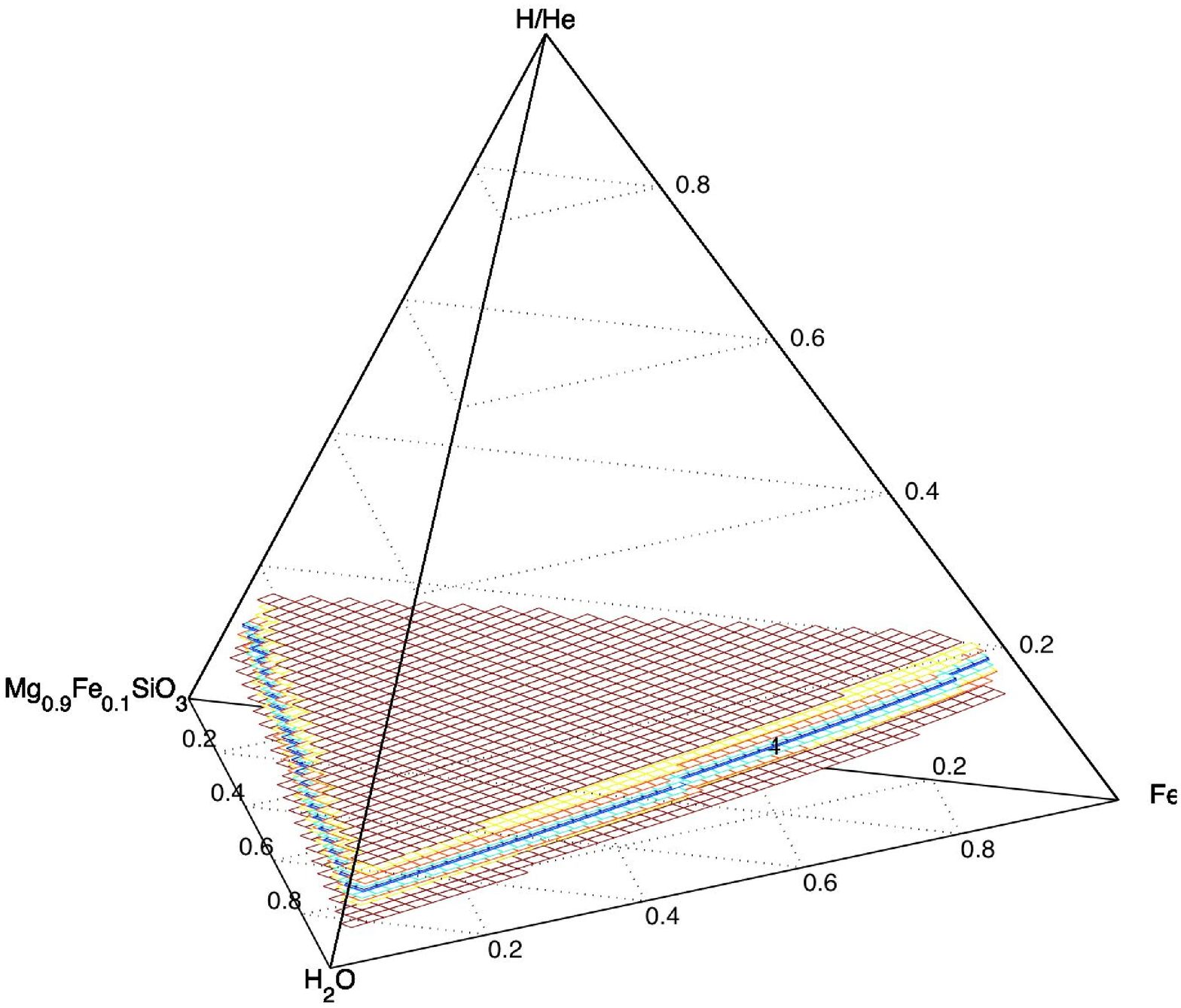}
\plotone{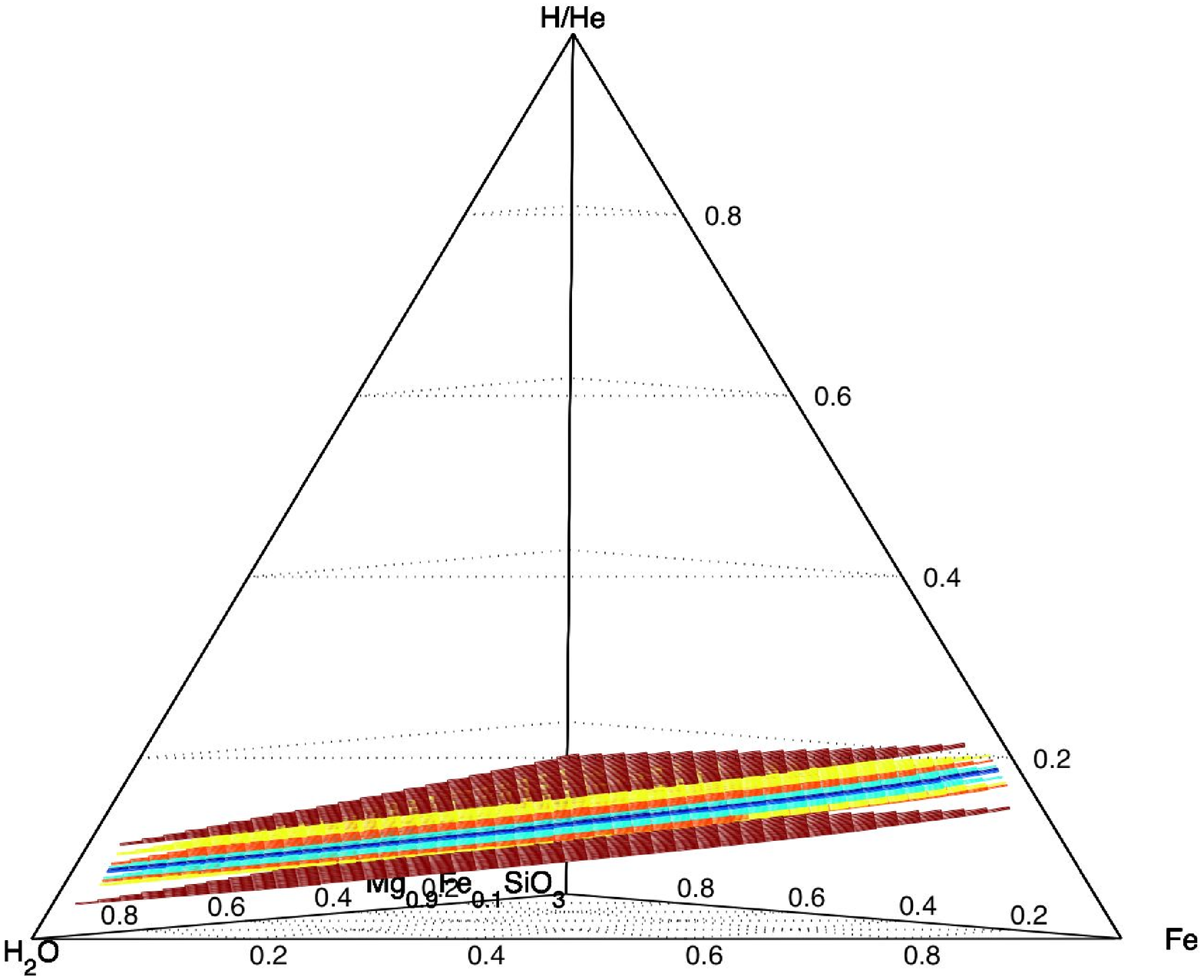}
\caption{Quaternary diagram for HAT-P-11b. Both model and observational uncertainties are taken into account to determine the plausible interior compositions of HAT-P-11b in this diagram. Two different views of the same quaternary diagram are shown. The surface of allowed compositions for our fiducial choice of model parameters ($M_p=25.8 M_{\oplus}$, $R_p=4.73 R_{\oplus}$, $T_0=867~\rm{K}$, $T_{\rm{eff}}=66~\rm{K}$, $\gamma=1$) is displayed in navy blue. To explore how uncertainties in model parameters weaken the constraints that can be placed on GJ~436b's interior composition, we vary each model parameter in turn while keeping all others fixed at their fiducial values. Two surfaces of the same color delimit the volume of composition space that is consistent with the range of values examined for each parameter. We consider $\gamma=0.1-10$ (cyan),  $T_{0}=778-1227~\rm{K}$ (green), and $T_{\rm{eff}}=58-86~\rm{K}$ (orange). The yellow surfaces denote the effect of varying the planet mass and radius within their $1\sigma$ observational uncertainties while maintaining all other model parameters at their fiducial values. Finally, the red surfaces delimit the full volume of possible compositions obtained by varying all parameters within the ranges described above. This figure is the HAT-P-11b analog to Figure~\ref{fig:GJ436bQuatParam} for GJ~436b.}
\label{fig:HATP11bQuatParam}
\end{figure}

Plausible interior compositions of HAT-P-11b are plotted in Figure~\ref{fig:HATP11bQuatParam}. Figure~\ref{fig:HATP11bQuatParam} displays the surface of allowed HAT-P-11b compositions for the fiducial parameter set ($M_p=25.8 M_{\oplus}$, $R_p=4.73 R_{\oplus}$, $T_0=867~\rm{K}$, $T_{\rm{eff}}=66~\rm{K}$, $\gamma=1$), and also shows the effect of considering a range of values for each model parameter. The range of values employed for each parameter ($\gamma=0.1-10$,  $T_{0}=778-1227~\rm{K}$, $T_{\rm{eff}}=58-86~\rm{K}$)  was determined following the procedure described in \S~\ref{sec:Ebudget}. The lower limit on the range of $T_{\rm{eff}}$ values considered had to be truncated at 58~K to avoid having the gas-layer \textit{P-T} profile enter an unphysical regime at high pressure and low temperatures (see \S~\ref{sec:Ebudget}). As for GJ~436b, uncertainties in the intrinsic luminosity of HAT-P-11b have an effect comparable to the $1~\sigma$ observational uncertainties, and significantly weaken the constraints we can place on the planet's interior composition.

We now attempt to compare the allowed compositions of HAT P-11 (Figure~\ref{fig:HATP11bQuatParam}) and GJ~436b (Figure~\ref{fig:GJ436bQuatParam}).  At  $\rho_p=1.33\pm0.20~\rm{g\,cm^{-3}}$~\citep{BakosEt2010ApJ}, HAT P-11b is less dense than GJ~436b ~\citep[$\rho_p=1.69^{+0.14}_{-0.12}~\rm{g\,cm^{-3}}$][]{TorresEt2008ApJ}. HAT P-11 could thus support a more massive gas layer (up to 19.0\%), and has a larger minimum gas mass fraction (7.1\%) for our fiducial choice of parameters. The effect of the average planet density on the gas layer constraints is partially mitigated by the higher level of stellar insolation received by HAT P-11b. When both $1~\sigma$ observational and model uncertainties are taken into account, the allowed compositions for HAT-P-11b and  GJ~436b overlap; it is plausible that HAT-P-11b and GJ~436b could both have the same proportion of core, mantle, water ices, and H/He gas layer.

Our comparison between the possible interior compositions of GJ~436b and HAT-P-11b is fraught with complications and should be interpreted with caution. Our conclusions contrasting the possible interior compositions of GJ~436b and HAT-P-11b are dependent on the method used to constrain the intrinsic luminosity of the hot Neptunes (see \S~\ref{sec:Ebudget}). Our constraints on the planets' internal heat flux  are admittedly rough and do not take into account the influence that two different levels of stellar irradiation could have on the luminosity evolution of these two planets. In addition, significant scatter in the observationally determined planetary masses and radii further hampers a comparative study of the transiting hot Neptunes' possible interior compositions. The GJ~436b radius obtained by \citet{BeanEt2008A&A} using \textit{HST} observations is larger than that found from  the infrared \textit{Spitzer} light curves with a $92\%$ formal significance, and would make GJ~436b less dense than HAT-P-11b. Improvements in the observational uncertainties on the mass and radii and in the constraints on the intrinsic luminosities of these two hot Neptunes are needed before we can truly make a robust comparison of their possible compositions. 

\section{Bayesian Inference Applied to Exoplanet Interior Structure Models}
\label{sec:Bayesian}

There are many model uncertainties that go into the interpretation of a measured mass and radius, and a major question is can we improve our deductions of interior composition from $M_p$ and $R_p$ by taking a more careful consideration of the uncertainties.  So far we have presented our planet interior composition constraints by delimiting a range of compositions on a ternary or quaternary diagram. In our presentation (Figures~\ref{fig:CoRoTnoH2OTern}, \ref{fig:Gl581dTern5}, \ref{fig:GJ436bQuatParam}, and \ref{fig:HATP11bQuatParam}), we know that it is more likely that the exoplanet's true composition falls within the $n~\sigma$ contours (surfaces) on the ternary (quaternary) diagram than outside the contours (surfaces). We do not know quantitatively, however, how likely it is that the exoplanet's true composition falls within the $n~\sigma$ bounds. In this section, we present an approach that yields a more detailed map of the relative likelihoods of the interior compositions on the ternary (quaternary) diagram and that takes all the contributing sources of uncertainty into account in a formal way. 

We turn to a more technical description of precisely what the contours in Figures~\ref{fig:CoRoTnoH2OTern}, \ref{fig:Gl581dTern5}, \ref{fig:GJ436bQuatParam}, and \ref{fig:HATP11bQuatParam} represent, and why there is a more thorough approach. The $n~\sigma$ contours (or surfaces in the case of quaternary diagrams) delimit the range of compositions that are consistent with the measured planetary mass  $\widehat{M_p}$ and radius $\widehat{R_p}$ (where the hats are used to distinguish measured values)  to within their $n~\sigma$ error bars for some choice of $\gamma$, $T_0$, and $T_{\rm{eff}}$ within the ranges described in \S~\ref{sec:Ebudget}. In other words, for every composition within the $n~\sigma$  shadings on the ternary or quaternary diagram, there is at least one choice of the model parameters within the parameter space cube $\left(\widehat{M_p}-n\sigma_{M_p}, \widehat{M_p}+n\sigma_{M_p}\right)\times\left(\widehat{R_p}-n\sigma_{R_p}, \widehat{R_p}+n\sigma_{R_p}\right)\times\left(\gamma_{min},\gamma_{max}\right)\times\left({T_{\rm{eff}}}_{min},{T_{\rm{eff}}}_{max}\right)\times\left({T_{0}}_{min},{T_{0}}_{max}\right)$ that yields a consistent solution. It is important to realize that the $n~\sigma$ contours in our ternary and quaternary diagrams do not represent confidence intervals. While one may make statements about the likelihood that the true planet mass and radius fall within $n~\sigma$ of their measured values, our $n~\sigma$ contours on the interior composition do not have a similar interpretation. This would be possible if only one model parameter were uncertain (for instance, if  $R_p$ had an observational uncertainty, while $M_p$ and all other model inputs were known exactly). In reality, however, there is more than one uncertainty (e.g., mass, radius, model inputs), and a more sophisticated technique is needed to draw accurate composition contours that can be associated with a likelihood. 

Bayesian statistics provide a more rigorous approach to calculate how different sources of uncertainty combine and translate into ambiguities on the interior composition of a planet. There are three categories of uncertainties. The first is observational uncertainties. The second is model uncertainties, in terms of the usually unconstrained range of input parameters (see \S~\ref{sec:Ebudget}). The third is the inherent degeneracy in interior compositions that yield a given mass and radius; in other words, the mapping from composition to mass and radius is not one-to-one. Using Bayesian statistics, we can associate every interior composition with a ``posterior likelihood", a number quantifying our degree of belief that the particular interior composition is the true interior composition of the planet (given our limited knowledge of the planet, and our assumptions). The ``posterior likelihood" function defined over the domain of possible interior mass distributions can then be used to draw well-defined contours (surfaces) in the ternary (quaternary) diagram for which the likelihood of the true composition falling within the contour can be stated. In \S~\ref{sec:results}, we are already drawing contours (surfaces) on ternary (quaternary) diagrams constraining the interior compositions of planets; Bayesian statistics provides an alternative way to accomplish this.

The foundation of Bayesian statistics is Bayes' Theorem, stated below in terms of the problem at hand (of inferring an exoplanets interior composition):

\begin{equation}
p\left(C|D,\,A\right)\propto\theta\left(C|A\right)\mathcal{L}\left(D|C,\,A\right).
\label{eq:BT}
\end{equation}

\noindent In the above expression, $C$ represents the set of all model parameters (including interior layer mass fractions, planet mass, planet radius, $\gamma$, etc.), $D$ represents all the measured data we have (measure planetary mass, planetary radius, stellar mass, stellar age, semimajor axis etc.), and $A$ denotes all of our assumptions (spherical symmetry, differentiated planet, negligible thermal corrections in the interior three layers, etc.). The function $\theta\left(C|A\right)$ is the prior probability of composition/parameters $C$ in the absence of measured data, given the assumptions. The priors $\theta$ incorporate assumptions about the range of model parameters to consider. They may also include detailed physics; for instance, one could assume a planet formation theory and use it to dictate a priori which interior compositions are more likely than others. Next, $\mathcal{L}\left(D|C,\,A\right)$ denotes the likelihood of the measured data $D$ for a given set of model parameters. Measurement uncertainties and correlations can be used to define the likelihood. Finally, $p\left(C|D,\,A\right)$ is the posterior likelihood of composition/model parameters $C$ given the measured data $D$ and the assumptions $A$. This is what we hope to calculate. The proportionality constant in Equation~(\ref{eq:BT}) is set so as to ensure that the posterior likelihood $p\left(C|D,\,A\right)$ is properly normalized. To make ternary diagram contour plots, one must marginalize (integrate) the posterior likelihood $p\left(C|D,\,A\right)$ over all model parameters (in the set $C$) other than the compositional layer mass fractions. The resulting marginalized posterior likelihood then represents the likelihood of a composition when the full range of nuisance parameter values is taken into account.

We now provide two examples to illustrate how the Bayesian approach described in the previous paragraph can be applied when drawing inferences about an exoplanet's interior. To begin, we apply Bayesian techniques to the case of a solid gas-less planet having measured mass and radius. We consider GJ~581d, adopting (as mentioned in \S~\ref{sec:GJ581d}) the GJ~581d minimum mass  $M_p=7.09~M_{\oplus}$ and two putative transit radii $R_p=1.5 R_{\oplus}$, and $R_p=2.0 R_{\oplus}$. We further assume that the measured planet mass and radius each have associated 5\% observational uncertainties. In what follows, we reproduce the GJ~581d composition constraints displayed in the Figure~\ref{fig:Gl581dTern5}  ternary diagrams, demonstrating how Bayesian statistics can be used to derive more informative and quantitative constraints on a transiting planet's interior composition.

 In this example, our assumptions $A$ include the following.
\begin{enumerate}
\item Our model described in \S~\ref{sec:model}  is appropriate to characterize the interior structure of GJ~581d.
\item GJ~581d does not have a significant gas layer.
\item GJ~581d has a pure iron core.
\item The Fe number fraction in the planet mantle is similar to that of the Earth ($\chi\approx0.1$).
\item The measurement uncertainties on the planetary mass and radius are Gaussian and uncorrelated.
\end{enumerate}

\begin{figure}
\epsscale{1.15}
\plotone{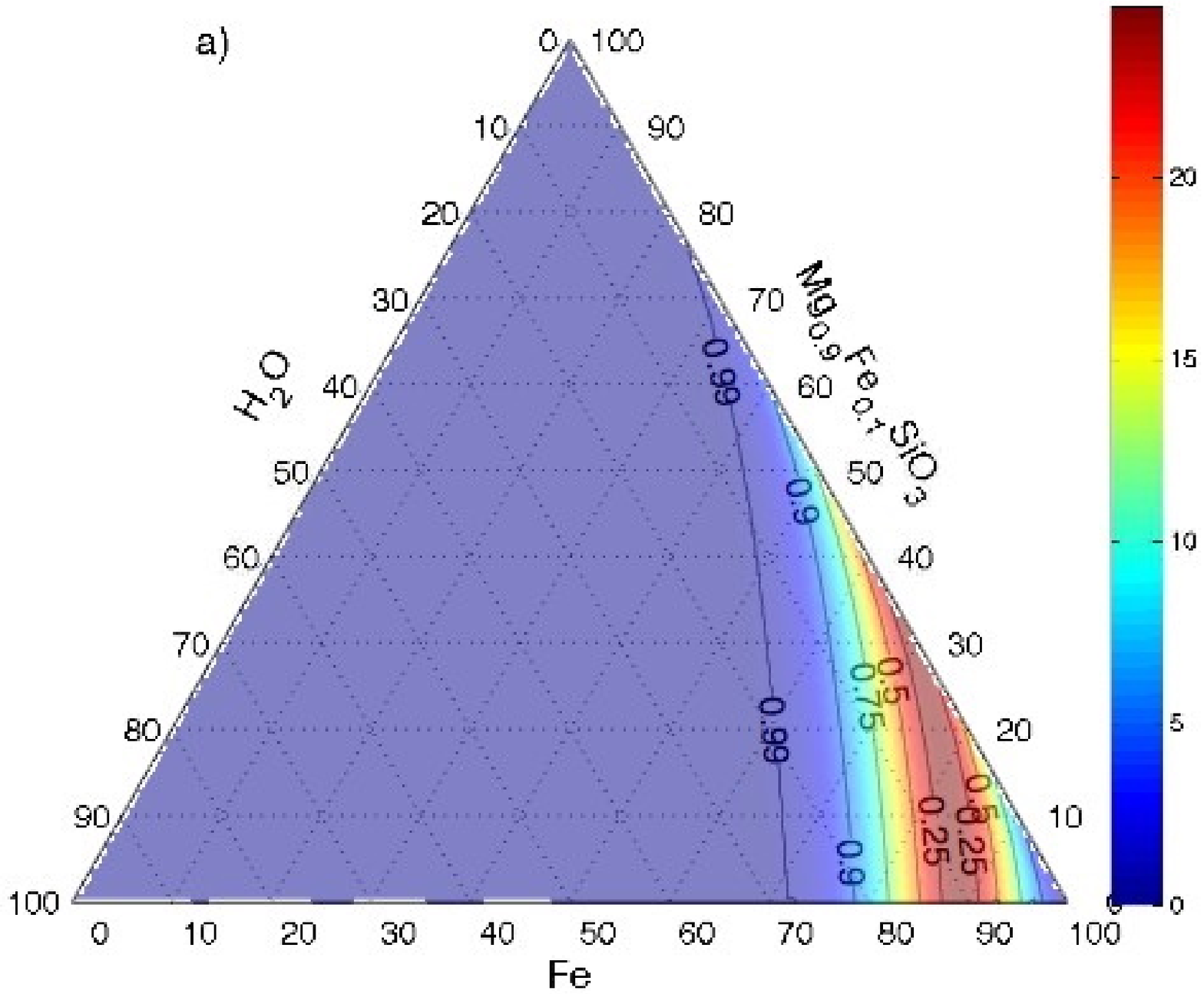}
\plotone{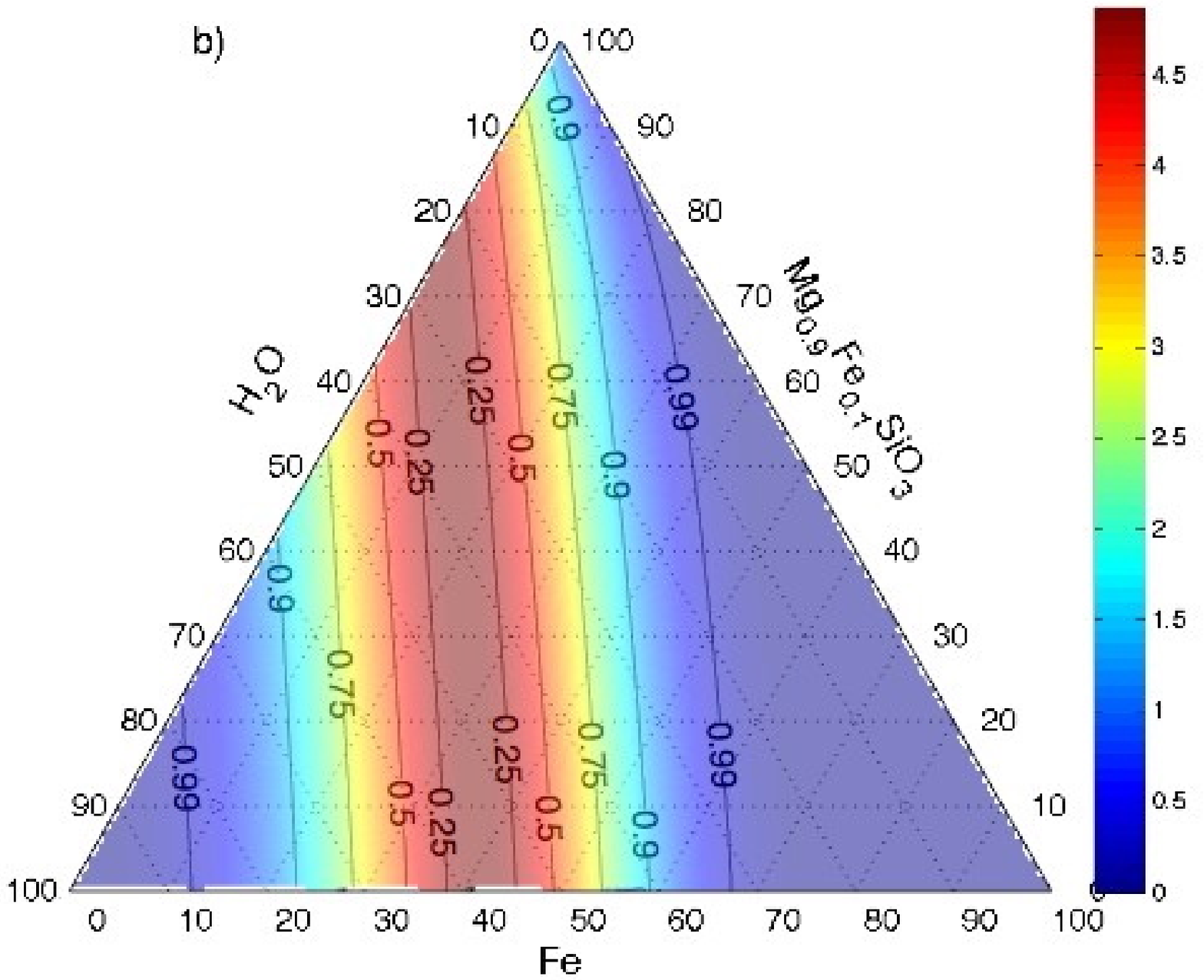}
\caption{GJ~581d interior composition posterior likelihood distribution. Only compositions without an H/He layer are considered. An observational uncertainty of 5\% is included on both the assumed mass  $\left(M_p=7.09~M_{\oplus}\right)$  and the assumed radii. Each diagram represents a different possible planetary radius: (a) $R_p=1.5 R_{\oplus}$ and (b) $R_p=2.0 R_{\oplus}$. The color shading in the ternary diagrams corresponds to the posterior likelihood distribution $p\left(x_{core}, x_{mantle}|D, A\right)$. Note that, for clarity, the two diagrams have different color scales. The contours are lines of constant posterior likelihood labeled with a Bayesian confidence value indicating the ``degree of belief" given the prior assumptions that the true composition of the planet falls within the contour. The confidence value is the integral of the posterior likelihood function over the surface within the contour. Compare these diagrams to Figures~\ref{fig:Gl581dTern5}(a) and (b), which show composition constraints obtained from the non-Bayesian approach employed in \S~\ref{sec:results} under assumptions identical to those used here.}
\label{fig:GJ581dBayes}
\end{figure}

\noindent Our model parameters in this case are $C\equiv\left(M_p, x_{core}, x_{mantle}\right)$, where $M_p$ is the planetary mass, and $x_i$ is the mass fraction in the \textit{i}th component. We do not explicitly include $x_{\rm{H_2O}}$ in the parameters  since it is determined by the constraint $1 =   x_{core} + x_{mantle} + x_{\rm{H_2O}}$. For a specified set of parameter values, our interior structure model will calculate a planetary radius $R_p\left(M_p, x_{core}, x_{mantle}\right)$. The data are the (putative) measured planetary mass and radius $D\equiv\left(\widehat{M_p}\pm\sigma_{M_p}, \widehat{R_p}\pm\sigma_{R_p}\right)$. We use the measured planetary mass and radius with their observational uncertainties to define the likelihood in terms of a Gaussian joint distribution for the planetary mass and radius

 \begin{eqnarray}
\lefteqn{\mathcal{L}\left(M_p, x_{core}, x_{mantle}|D, A\right)}\nonumber\\
&&=\frac{1}{2\pi\sigma_{M_p}\sigma_{R_p}} e^{-\left(M_p-\widehat{M_p}\right)^2/2{\sigma_{M_p}}^2-\left(R_p-\widehat{R_p}\right)^2/2{\sigma_{R_p}}^2},\nonumber\\
\label{eq:like}
\end{eqnarray}

\noindent where $R_p\equiv R_p\left(M_p, x_{core}, x_{mantle}\right)$. In this example, we take a flat prior $\theta\left(M_p, x_{core}, x_{mantle}|A\right)\propto1$, for which regions of composition space having equal area on the ternary diagram are equally likely. This prior is analogous to what we implicitly assumed when plotting the $n\sigma$ contours in Figure~\ref{fig:Gl581dTern5}. Given the assumed prior  $\theta\left(M_p, x_{core}, x_{mantle}|A\right)$, we multiply $\theta$ and $\mathcal{L}$ to obtain the posterior likelihood $p\left(M_p, x_{core}, x_{mantle}|D, A\right)$ following Equation~(\ref{eq:BT}). We then marginalize over the planetary mass $M_p$, obtaining a posterior likelihood depending only on the interior composition, 
 \begin{eqnarray}
\lefteqn{p\left(x_{core}, x_{mantle}|D, A\right)}\nonumber\\
&&=\int_{0}^{\infty}p\left(M_p, x_{core}, x_{mantle}|D, A\right)\rm{d}M_p, 
\end{eqnarray}

\noindent for plotting on a ternary diagram (Figure~\ref{fig:GJ581dBayes}). In Figure~\ref{fig:GJ581dBayes} we show contours of constant posterior likelihood, and label each contour with the posterior likelihood that the true composition lies inside the contour (calculated by integrating the posterior likelihood function over the area within the contour). When drawn in this way, the composition contours in the ternary diagrams are Bayesian confidence regions with confidence values that should be interpreted as the ``degree of our belief" that the true composition of a planet falls within the contour given our assumptions and our observations of the planet. 

Applying a Bayesian analysis to the putative mass and radius measurements of GJ~581d, we extract more informative and quantitative composition constraints than those obtained from the non-Bayesian analysis in \S~\ref{sec:GJ581d}. The non-Bayesian $n\sigma$ contours in Figure~\ref{fig:Gl581dTern5} effectively denote the loci of interior compositions for the discrete mass-radius pairs $\left(\widehat{M_p}\pm n\sigma_{M_p}, \widehat{R_p}\mp n\sigma_{R_p}\right)$. By contrast, the results of our Bayesian analysis (shown in Figure~\ref{fig:Gl581dTern5}) take into account the full mass-radius relationship for each possible interior composition. While Figure~\ref{fig:Gl581dTern5} does not give any indication of the relative plausibility of two different compositions within the same $n\sigma$ contour, the Bayesian framework yields a posterior likelihood map $p\left(x_{core}, x_{mantle}|D, A\right)$ over the entire ternary diagram (shown by the color shading in Figure~\ref{fig:GJ581dBayes}). Finally, on its own, Figure~\ref{fig:Gl581dTern5} does not reveal an estimate of how likely it is that the true composition of the GJ~581d falls within its $n\sigma$ bounds. The contours in Figure~\ref{fig:GJ581dBayes} are, however, associated with Bayesian confidence values. Comparing Figures~\ref{fig:Gl581dTern5} to \ref{fig:GJ581dBayes} we see that in this case, given our assumptions, the true composition of the GJ~581d should fall within the $1\sigma$ contours in Figure~\ref{fig:Gl581dTern5} with a Bayesian confidence of roughly 75\%. For the price of having to assume a prior $\theta\left(C|A\right)$, Bayesian inference yields more detailed and quantitative constraints on a transiting exoplanet's composition than other analysis approaches. 

We now present a second example to demonstrate the effect of priors. We consider CoRoT-7b, and for illustration we make several different assumptions about the iron concentration in its mantle. As mentioned in \S~\ref{sec:CoRoT7b}, we assume that CoRoT-7b has a pure iron core and does not have a significant water or gas layer. With these restrictions, our model parameters are $C\equiv\left(M_p, x_{core}, x_{\rm{FeSiO_3}}\right)$, where $x_{\rm{FeSiO_3}}$ and $x_{\rm{MgSiO_3}}$ denote the fraction of the planet's mass consisting of mantle $\rm{FeSiO_3}$ and mantle $\rm{MgSiO_3}$ respectively $\left(x_{\rm{MgSiO_3}} = 1 - x_{core}-x_{\rm{FeSiO_3}}\right)$. We proceed to calculate the interior composition posterior likelihood function $p\left(x_{core}, x_{\rm{FeSiO_3}}|D, A\right)$ following an analogous procedure to that outlined in detail in the GJ~581d example above. Again assuming that the measurement uncertainties on the planetary mass and radius are Gaussian and uncorrelated, we define the likelihood in terms of a Gaussian joint distribution for the planetary mass and radius, as given in Equation~(\ref{eq:like}). Our choices for the priors $\theta\left(M_p, x_{core}, x_{\rm{FeSiO_3}}|A\right)$ are described below.

\begin{figure*}
\epsscale{1.15}
\plotone{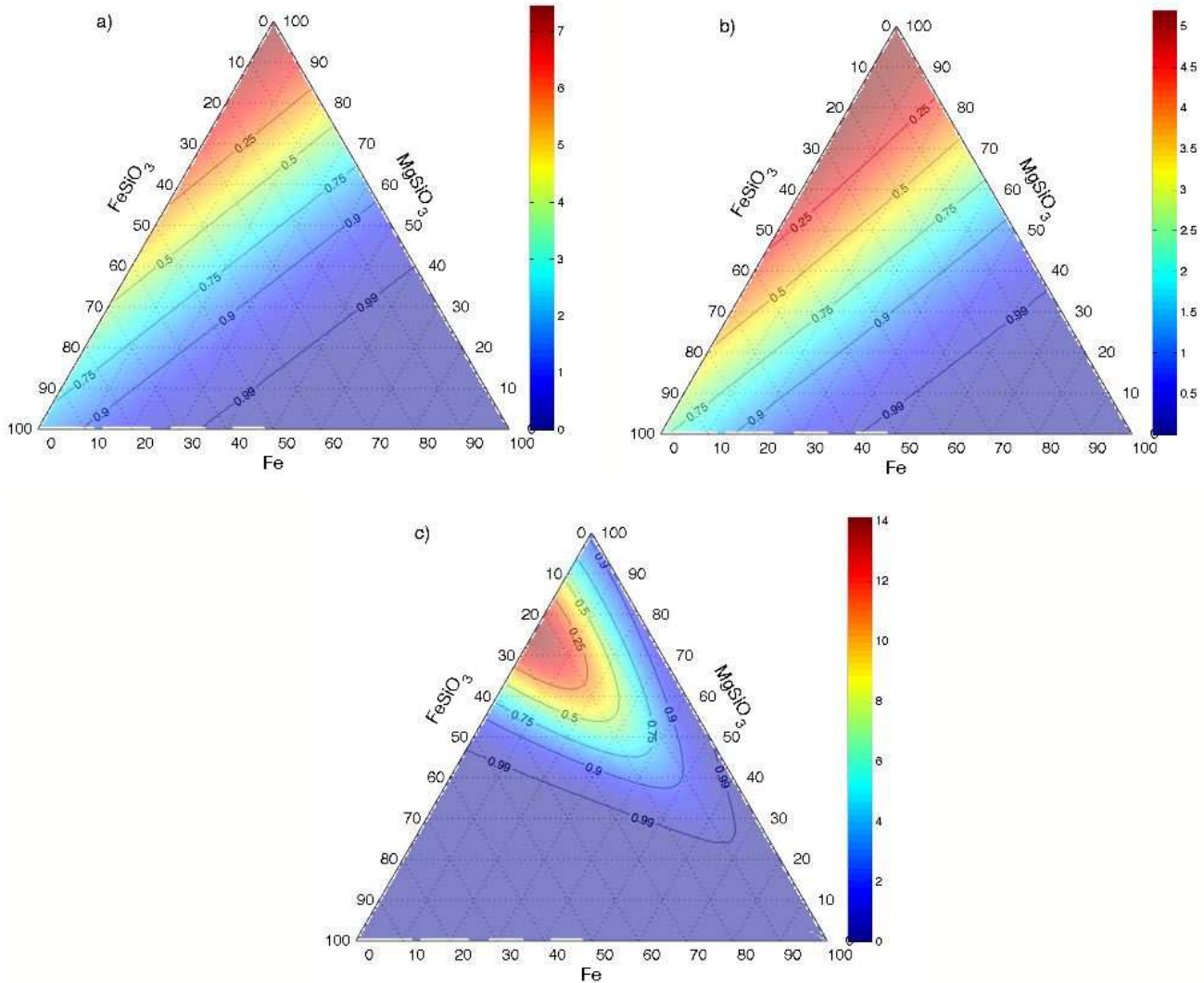}
\caption{CoRoT-7b interior composition posterior likelihood distribution. Three different choices of prior are shown. In panel (a) we take a flat prior on the core and FeSiO$_3$ mass fractions, $\theta_1\left(M_p, x_{core}, x_{\rm{FeSiO_3}}|A\right)\propto 1$. In panel (b) we adopt a uniform prior on the mantle iron number fraction, $\chi$ (Equation~(\ref{eq:theta2})). Finally, in panel (\textit{c}) we take a strong prior on the mantle iron number fraction, in which $\chi$ is assumed to have a Gaussian distribution with mean $\bar\chi=0.1$ and standard deviation $\sigma_{\chi}=0.1$ (Equation~(\ref{eq:theta3})).  The color shading in the ternary diagrams corresponds to the posterior likelihood distribution $p\left(x_{core}, x_{\rm{FeSiO_3}}|D, A\right)$. Note that, for clarity, each diagram has a different color scale. The contours are lines of constant posterior likelihood labeled with a Bayesian confidence value indicating the ``degree of belief" given the prior assumptions that the true composition of the planet falls within the contour. The confidence value is the integral of the posterior likelihood function over the surface within the contour.}
\label{fig:CoRoT7bBayes}
\end{figure*}

	For illustration, we consider three different choices for the prior $\theta\left(M_p, x_{core}, x_{\rm{FeSiO_3}}|A\right)$ in Figure~\ref{fig:CoRoT7bBayes}.  In all three cases, we take $\theta$ to be independent of $M_p$ so that mass intervals of equal size $dM_p$ are equally likely (before taking into account radial velocity observations measuring CoRoT-7b's mass).

	For our first prior (Figure~\ref{fig:CoRoT7bBayes}(a)), we take a flat prior in which each division of mass between the Fe core, MgSiO$_3$, and FeSiO$_3$ is equally likely: $\theta_1\left(M_p, x_{core}, x_{\rm{FeSiO_3}}|A\right)\propto 1$. This prior is most similar to what we have implicitly assumed in plotting Figure~\ref{fig:CoRoTnoH2OTern} and corresponds to the case in which regions of equal area on the ternary diagram are, a priori, equally likely. 
		
	Second, in Figure~\ref{fig:CoRoT7bBayes}(b) we choose a prior where the mantle iron number fraction, $\chi$, is uniformly distributed between 0 and 1 (with all possible values $x_{core}$ also equally likely). The Fe number fraction $\chi$ of a silicate material is defined as the ratio of $\rm{Fe/(Mg+Fe)}$ by number. For perovskite $\rm{Mg_{1-\chi}Fe_{\chi}SiO_3}$, $\chi$ is related to the mass fractions $x_{\rm{MgSiO_3}}$ and $x_{\rm{FeSiO_3}}$ through
\begin{eqnarray}
\lefteqn{\chi\left(x_{core}, x_{\rm{FeSiO_3}}\right)} \nonumber\\
&&= \frac{x_{\rm{FeSiO_3}}}{x_{\rm{FeSiO_3}}+ \left(1-x_{core}-x_{\rm{FeSiO_3}}\right)\left(\mu_{\rm{FeSiO_3}}/\mu_{\rm{MgSiO_3}}\right)},\nonumber\\
\end{eqnarray}
\noindent where $\mu_{\rm{MgSiO_3}}$ and $\mu_{\rm{FeSiO_3}}$ are the molar weights of MgSiO$_3$ and FeSiO$_3$ respectively. Transforming from a uniform prior in $x_{core}$ and $\chi$ to the variables $x_{core}$ and $x_{\rm{FeSiO_3}}$ we find 
\begin{eqnarray}
\lefteqn{\theta_2\left(M_p, x_{core}, x_{\rm{FeSiO_3}}|A\right)\propto\left|\frac{\partial\chi}{\partial x_{\rm{FeSiO_3}}}\right|}\nonumber\\
&&=\frac{\left(1-x_{core}\right)\left(\mu_{\rm{FeSiO_3}}/\mu_{\rm{MgSiO_3}}\right)}{\left(x_{\rm{FeSiO_3}} + \left(1-x_{core}-x_{\rm{FeSiO_3}}\right)\left(\mu_{\rm{FeSiO_3}}/\mu_{\rm{MgSiO_3}}\right)\right)^2},\nonumber\\
\label{eq:theta2}
\end{eqnarray}	
\noindent where the right-hand side of Equation~(\ref{eq:theta2}) is the Jacobian determinant of the transformation. It is important to note that assuming a uniform prior probability on $x_{\rm{FeSiO_3}}$ is not the same thing as assuming a uniform prior probability on $\chi$ (or in other words $\theta_1\neq\theta_2$). A uniform prior probability on $\chi$ effectively weights compositions having small mantle mass fractions more heavily those having large mantle mass fractions. This is because when $x_{\rm{FeSiO_3}}+x_{\rm{MgSiO_3}}=1-x_{core}$ is small, small increments in $x_{\rm{FeSiO_3}}$ can correspond to large changes in $\chi$.

For our final prior (Figure~\ref{fig:CoRoT7bBayes}\textit{c}), we take an extreme case for illustration, adopting a strong prior. We assume that CoRoT-7b's mantle has an iron fraction similar to that of the Earth. Specifically, we take $\chi$ to have a Gaussian distribution with mean $\bar\chi=0.1$ and standard deviation $\sigma_{\chi}=0.1$:
\begin{eqnarray}
\lefteqn{\theta_3\left(M_p, x_{core}, x_{\rm{FeSiO_3}}|A\right)}\nonumber\\
&&\propto\frac{\left(1-x_{core}\right)\left(\mu_{\rm{FeSiO_3}}/\mu_{\rm{MgSiO_3}}\right)e^{-\left(\chi-\bar\chi\right)^2/2\sigma^2_{\chi}}}{\left(x_{\rm{FeSiO_3}} + \left(1-x_{core}-x_{\rm{FeSiO_3}}\right)\left(\mu_{\rm{FeSiO_3}}/\mu_{\rm{MgSiO_3}}\right)\right)^2}.
\label{eq:theta3}\nonumber\\
\end{eqnarray}	
\noindent In Equation~(\ref{eq:theta3}), $\chi\equiv\chi\left(x_{core}, x_{\rm{FeSiO_3}}\right)$. Our third choice of prior has a strong effect on the posterior likelihood distribution in Figure~\ref{fig:CoRoT7bBayes}\textit{c}, favoring compositions near the $\chi=0.1$ line in the ternary diagram.

The danger in the Bayesian approach described in this section is that one's prior assumptions will affect the compositional likelihoods, as illustrated by the CoRoT-7b example above. While modelers have not formally been using the Bayesian approach, they have been making critical assumptions that affect their interior composition interpretation of the mass and radius. For example, \citet{Zeng&Seager2008PASP} assumed a uniform distribution of allowed compositions; \citet{ValenciaEt2007bApJ} excluded certain regions of the ternary diagram having low mantle mass fractions, but effectively presented all remaining compositions as equally likely; and \citet{FigueiraEt2009A&A} used a planet formation and migration model to predict which bulk compositions of GJ~436b may be more likely. The Bayesian approach above provides a framework in which the priors are explicitly stated, whether they are flat or not. In this way, the effect on the results of choosing different prior assumptions can be quantified. 

Bayesian inference may or may not be the best approach to interpret an exoplanet's measured mass and radius. Our goal was to take into account, in a formal way, all the sources of uncertainty contributing to ambiguities in a planet's interior composition. We have shown that the Bayesian approach is a way to meet this goal. Less formal approaches (such as that in \S~\ref{sec:results}) for constraining a planet's interior composition can also be insightful, but one should be heedful of how multiple sources of uncertainty are combined together when interpreting their composition bounds. A problem with the Bayesian approach is that, since the data available on any given transiting planet are limited, the priors assumed can have an important effect on the results. As long as the effect of the priors is explored and acknowledged, Bayesian statistics can help to maximize the compositional inferences we can draw from the limited data that we have on distant exoplanets. Regardless of any statistical approach taken, modelers must be explicit about their prior assumptions and about the precise significance of their compositional constraints.

\section{Discussion}
\label{sec:Dis}

\subsection{External Constraints on Planetary Composition}

So far, we have only considered the constraints placed on an exoplanet's interior composition by mass and radius measurements alone. In this section, we discuss how planetary formation theories, compositional stability, and cosmic elemental abundances  can be used to place additional constraints on a planet's interior composition.

Planet formation models predict that some interior compositions are more likely to form than others. \citet{ValenciaEt2007bApJ} considered the constraints imposed by protoplanetary disk abundances, adopting the point that planets with large iron cores or large water ice layers but small silicate mantles are very unlikely. From the relative abundance of Si and Fe in the solar nebula ($\rm{Si/Fe}\sim0.6$), \citet{ValenciaEt2007bApJ} propose a minimum  ratio of mantle to core mass. Further, \citet{ValenciaEt2007bApJ} put forward that, since comets are dirty snowballs comprised of both volatiles (water) and dust, cometary delivery of water to a planet will simultaneously deliver silicates to build up the planet's mantle at the rate of at least $\rm{Si/H_2O}\sim0.23$ by mass. \citet{GrassetEt2009ApJ} 
choose a distribution of Mg/Si and Fe/Si molar ratios in the bulk compositions of their modeled planets based on the measured abundances in a collection of planet-hosting stars. Our approach is to consider the full ternary diagram and to avoid imposing strong priors on the a priori relative likelihood of various interior compositions. In this way, we limit the effect of planet formation assumptions on the composition constraints we derive.

Planet formation models also constrain the mass and composition of hydrogen and helium gas layers.
In this paper, we have only considered a fixed ratio of H/He. Planet atmospheres may have a range different from solar, based on the atmosphere formation process. 
Planet atmospheres can originate from capture of nebular gases, degassing during accretion, and degassing during subsequent tectonic activity. Outgassing would produce a hydrogen atmosphere with negligible helium, because helium is not trapped in rocks \citep{HeberEt2007GeCoA}. The mass of the atmosphere created from outgassing, however, could have a wide range \citep{ElkinsTanton&Seager2008aApJ}. In contrast, the H/He composition of a gas layer captured from a nebula would reflect the composition of the nebula (modulated by ensuing atmospheric escape) and is presumably close to solar. It is conventional to accept that accretion of nebula gases is most important for massive protoplanetary cores; accretion of nebular gas is expected for rocky cores above $10~M_{\oplus}$ while often neglected for planets below  $6~M_{\oplus}$ \citep[e.g.][]{SelsisEt2007Icarus}. We show the full quaternary diagram because planets in the intermediate-mass range $6-10~M_{\oplus}$ (such as GJ~581d) may still accrete a significant mass of H-rich gas \citep{AlibertEt2006A&A, Rafikov2006ApJ} and retain it under the right conditions. 

A natural question in exploring the interior composition range of a hot super-Earth is whether or not a hot super-Earth can retain an interior water layer. We can set upper limits on the rate at which the low-mass exoplanet CoRoT-7b $\left(M_p=4.8\pm0.8~M_{\oplus}\right)$ would lose H$_2$O. CoRoT-7b is extremely close to its host star and suffering intense irradiation; the surface temperature is 1800-2600~K at the sub-stellar point depending on the planet Albedo and energy redistribution \citep{LegerEt2009A&A}. Scaling the results of \citet{SelsisEt2007Icarus} to CoRoT-7b's semimajor axis and host star luminosity, we find a minimum water content lifetime of 0.07-1.0~Gyr.  \citet{SelsisEt2007Icarus} set upper bounds on the water mass loss of ocean planets around Sun-like stars, considering both energy-limited thermal escape driven by extreme UV and X-ray irradiation heating the planet exospheres, and non-thermal escape driven by erosion from the stellar wind. Although we do not have an upper bound on the water content lifetime, we do not consider the presence of a water layer (or of H/He) on CoRoT-7b because of its extreme proximity to the host star. If CoRoT-7b does in fact have a significant water content, it would be in the form of a super-fluid H$_2$O envelope with no liquid-gas interface \citep[see, e.g.,][]{LegerEt2004Icarus, SelsisEt2007Icarus}. \citet{ValenciaEt2009AstroPh} considered the possibility that CoRoT-7b might harbor a water vapor layer.

Atmospheric escape is another process that is difficult to model but could potentially be helpful in interpreting the composition of a planet by ruling out regions of the quaternary diagram. It is difficult to predict atmospheric escape as it depends on the detailed physical characteristics of the planet's atmosphere and its interaction with the stellar insolation. Examples of properties on which atmospheric escape rates depend are the composition of the atmosphere, the thermal structure of the atmosphere, the UV history of the host star, the density of the stellar wind, the speed of the stellar wind, and the planet's intrinsic magnetic moment. In order to understand whether or not a planet has retained any hydrogen, one would have to model a specific exoplanet, taking into consideration the range of possibilities for the factors controlling atmospheric escape. As an approximation, \citet{LecavelierDesEtangs2007A&A} has considered energy-limited atmospheric escape to estimate atmosphere lifetimes. Following his approach, we estimate escape rates of $3\times10^7~\rm{kg\,s^{-1}}$  and $5\times10^7~\rm{kg\,s^{-1}}$ for GJ~436b and HAT-P-11b, respectively. As a second example, \citet{SelsisEt2007A&A} have found that at GJ~581A's current X-ray and EUV luminosity, GJ~581d should not currently be experiencing extreme atmospheric mass loss, although atmospheric erosion rates at earlier (and more active) stages in the GJ~581 system's lifetime are uncertain. It is fair to say that no published models conclusively detail the mass-loss history of a given Neptune or super-Earth exoplanet, and it is not clear for which cases this is even possible.

\subsection{Chemical Composition of Interior Layers}

The chemical make-up of a transiting planet's envelope, ices, mantle, and core is not known a priori. In this work, we have limited the chemical compositions that we consider for the interior layers of a planet to an H/He gas layer with solar composition, water ices, perovskite mantle, and a predominantly iron core. We have explored the effect of varying the mantle iron content and of including a light element in the planet core (see \S~\ref{sec:results}). We selected our fiducial choice for the chemical compounds comprising the interior planetary layers in our model to capture the dominant materials making up the solar system planets. 

There are, however, several additional possibilities for the chemical make up of an exoplanet. For example, ammonia ices change the EOS of NeptuneÕs interior \citep[e.g.][]{PodolakEt1995P&SS}. Super-Earths that have outgassed an extended hydrogen envelope would lack helium \citep{ElkinsTanton&Seager2008aApJ}. A massive CO gas layer in a hydrogen-poor planet would have a different EOS than an H/He gas layer, but because of its density, likely would not contribute to an extended radius. Water-dominated ``ocean" planets could have a vapor atmosphere or even a superfluid surface layer \citep[e.g.,][]{Kuchner2003ApJ, LegerEt2004Icarus, SelsisEt2007Icarus}. Carbon planets will have different interior compositions entirely, as compared to silicate-based planets \citep{Kuchner&Seager2005AstroPh, SeagerEt2007ApJ}. When interpreting the mass and radius for a given exoplanet, these other compositions should be included in the future. 

\subsection{Planet  Evolution}
\label{sec:disevolution}

We have found that uncertainties in an H/He-laden planet's intrinsic luminosity significantly weaken the constraints that can be placed on the planet's interior composition. In the case of GJ~436b and HAT-P-11b, uncertainties in $T_{\rm{eff}}$ even dominate the observational uncertainties on the planet masses and radii. Developing models to better predict a low-mass planet's intrinsic luminosity is thus an important endeavor to further our ability to study the interior compositions of super-Earths and hot Neptunes.

Time-dependent simulations of planets as they age and cool could be employed to constrain the planets' intrinsic luminosities. Exoplanet evolution calculations have been performed in several previous studies \citep[e.g.][]{BaraffeEt2003A&A, BurrowsEt2003ApJ, ChabrierEt2004ApJ, Fortney&Hubbard2004ApJ, BaraffeEt2006A&A, FortneyEt2007ApJ, BaraffeEt2008A&A}. Simulations of the solar system giants have illustrated how complicated the process of predicting a planet's intrinsic luminosity can be. While simple models of Jupiter's evolution and interior structure are in good agreement with the observed cooling rate \citep{Hubbard1977Icarus}, homogeneous contraction models predict intrinsic luminosities that are too low for Saturn \citep{Guillot2005AREPS} and too high for Uranus and Neptune \citep{Stevenson1982P&SS}. 

 In this work, we have subsumed an evolution calculation by using a simple scaling relation to derive a plausible intrinsic luminosity range from a planet's mass, radius, and age. An evolution calculation coupled with our planet interior model may eventually offer a more self-consistent approach to constrain the intrinsic luminosities of low-mass exoplanets. The addition of a time-dependent cooling calculation would essentially shuffle our uncertainties in $T_{\rm{eff}}$ to uncertainties in temperature-dependent EOSs, the planet's chemical composition, and the planet's migration, geological, tidal evolution, and compositional histories. Poorly constrained planet ages further limit the improvements an evolution calculation could provide in the intrinsic luminosity constraints.  These limitations and the added computational power demanded by time-dependent models motivate our use of an approximate phenomenological approach to constrain $T_{\rm{eff}}$. In a future paper, we plan to perform an evolution calculation to verify that the range of $T_{\rm{eff}}$ values chosen in this work is representative of the uncertainties in a planet's age and history. 
 
 \subsection{Beyond Mass and Radius - Further Observational Constraints on Compositions}

In this paper, we have focused on the constraints that can be placed on a transiting exoplanet's interior composition using only knowledge about its mass, radius, and stellar insolation (all properties that can be measured by current spectroscopic or photometric techniques).  Are there other observations that can further restrict the range of interior compositions of a low-mass exoplanet? Transmission spectra during primary transit can be used to discriminate between a planet with a significant hydrogen envelope and a hydrogen-poor super-Earth \citep{MillerRicciEt2009ApJ}. Observations will be extremely challenging, even with multiple transits \citep{Kaltenegger&Traub2009ApJ, DemingEt2009PASP}. In the case of close-in transiting hot Jupiters, apsidal precession induced by tidal bulges on the planet could produce observable changes in the transit light-curve shape, revealing additional information about the interior density distribution of the planet \citep[through the Love number;][]{Ragozzine&Wolf2009ApJ}. This idea is geared at hot Jupiters and it is unclear whether the effect will be significant for terrestrial or Neptune-size planets. The potential for an improved understanding of planetary interiors should provide strong motivation for the advancement of these observational techniques toward greater sensitivities. 

\section{Conclusions}
\label{sec:con}

We have quantified how observational uncertainties, model uncertainties, and inherent degeneracies all contribute to the range of plausible bulk compositions for transiting low-mass exoplanets. We have only considered the constraints imposed on the composition by the measured planetary mass, radius, and stellar insolation, and did not speculate on the formation history. Uncertainties in the formation, evolution, and age of the planets studied were encapsulated in the range of values chosen for the internal heat flux, albedo, $\gamma$, and mantle iron content. We summarize our main conclusions below. 

\begin{enumerate}

\item The interior compositions of CoRoT-7b, GJ~436b, and HAT-P-11b (the three lowest mass transiting planets known to date) are constrained by our interior structure model.  

\begin{itemize}
\item CoRoT-7b: An Earth-like composition having $30\%$ of its mass in an iron core and the remaining $70\%$ of its mass in a silicate mantle is consistent with the measured mass and radius within $1~\sigma$. Large core mass fractions  ($x_{\rm{Fe}}\gtrsim0.76-0.86$) are ruled out at the level of at least $3~\sigma$, but all other combinations of core mass fraction and iron mantle content (in water-less, gas-less compositions) are allowed; the planet could have no core or could be composed of up to 86\% pure iron by mass and still fall within the $3~\sigma$ error bars on $M_p$ and $R_p$. If CoRoT-7b does not contain significant amounts of water or gas, some of the mass-radius pairs within $M_p\pm1\sigma_M$ and $R_p\pm1\sigma_R$ (specifically those that correspond to bulk densities lower than a pure silicate planet) can be ruled out.

\item GJ~436b: GJ~436b must have between 2.3\% and 15.5\% H/He layer by mass to produce the observed transit depth. These lower and upper limits on the GJ~436b H/He layer depend on the intrinsic luminosity of the planet. The water content of GJ~436b is very poorly constrained by the mass and radius measurements alone: GJ~436b could be completely dry, or could alternatively consist of up to 96.4\% water by mass. 

\item HAT-P-11b: Nominally, HAT-P-11b's  measured density is lower than GJ~436b's. HAT-P-11b thus requires a higher minimum mass of gas  (at least 7.1\%) and can support a more massive gas envelope (up to 19.0\% by mass). Comparisons between the range of plausible compositions for GJ~436b and HAT-P-11b are made difficult by the uncertain intrinsic luminosities of these planets and by the scatter in the observationally determined masses and radii for each planet.
\end{itemize}

\item Uncertainties in the intrinsic luminosities of low-mass exoplanets significantly weaken the compositional constraints that can be derived from a pair of mass and radius measurements. In the case of both GJ~436b and HAT-P-11b, the uncertainties on $T_{\rm{eff}}$ dominate the observational uncertainties. Better constraints on $T_{\rm{eff}}$ (possibly obtained through planetary evolution models) are required to improve our limits on the interior compositions of transiting hot Neptunes.

\item The degree to which we can constrain the composition of a super-Earth depends on the planet's density. Putative planets with extreme densities (especially those with very high densities) allow the tightest composition constraints (assuming similar observational uncertainties on $M_p$ and $R_p$). Denser planets will have smaller radii, however, making it more difficult to measure their transit radii with high precision. 

\item Quaternary diagrams provide a convenient way to illustrate the range of possible interior compositions for a transiting planet that harbors a significant gas layer. They allow one to display interior compositions consisting of four distinct components (in our case an iron core, silicate mantle, water ice layer, and H/He envelope). 

\item When constraining the interior compositions of transiting exoplanets, modelers must include in their analysis many sources of uncertainty (model, observational, and inherent degeneracy). The Bayesian approach presented in \S~\ref{sec:Bayesian} provides a framework with which one can combine all the sources of uncertainty contributing to ambiguities in a planet's interior composition in a formal way. Given explicitly stated assumptions and the measured planet parameters, the procedure described in \S~\ref{sec:Bayesian} outlines how to calculate the relative likelihood that any interior composition on the ternary diagram is the true composition of the planet. The likelihoods obtained can be strongly dependent on the prior assumptions made.  In the Bayesian framework, however, the prior assumptions are explicitly stated and so their effect can be explored and quantified. 

\item Allowing for the possibility of a gas layer in future interpretations of the mass and radius measurements  of transiting super-Earths will greatly increase the range of possible interior compositions of the planet. The presence of even a low-mass gas layer contributing to the transit radius can significantly alter the inferred characteristics of the underlying solid planet. Specifically, the higher the gas mass fraction the denser the solid planet interior must be to compensate. 

\end{enumerate}

Planetary science has come a long way toward understanding planetary interiors.  With Jupiter, we know that its bulk composition is dominated by 50\%-70\% hydrogen by mass; that the helium abundance in its atmosphere is somewhat below the protosolar value; that it contains  $1~M_{\oplus}\lesssim M_Z\lesssim 39~M_{\oplus}$ of heavy elements; and that between 0 and 11~$M_{\oplus}$ of the heavy elements could be concentrated in a core~\citep{Saumon&Guillot2004ApJ}. For exoplanets, without recourse to in situ composition measurements and gravitational moment measurements from spacecraft flybys we will be permanently limited in what we can infer about the interior composition from the measured mass and radius. Not only are the measurement uncertainties relatively large (2\% at best on $R_p$ compared to $\sim 0.01\%$ for solar system planets), but models are also needed to map the planetary mass and radius into interior composition and the model uncertainties are high. We will have to be satisfied simply knowing that we can quantify the wide range of exoplanet plausible interior compositions.

\acknowledgments

We thank Thomas Beatty for useful comments and discussions. We also thank Diana Valencia, whose referee comments helped to strengthen this paper. This work is supported by the Natural Sciences and Engineering Research Council of Canada.

\bibliography{exoplanets}

\clearpage

\end{document}